%% file: cTL_MDM.tex
\documentclass[aps,prb,reprint,twocolumn,superscriptaddress,showpacs,showkeys,a4paper]{revtex4}

\usepackage{amsmath,amssymb,amstext}
\usepackage[usenames,dvipsnames]{color}
\usepackage{graphicx,subfigure,relsize}
\usepackage{bm,bbold,braket}
\usepackage{natbib}

\usepackage{amsmath}
\usepackage{amsfonts}
\usepackage{amssymb}
\usepackage{verbatim}
\usepackage{color}
\usepackage{epsfig}
\usepackage{subfigure}
\usepackage{exscale}
\usepackage{float}

\begin{document} 

\title{Cluster Luttinger liquids and emergent supersymmetric conformal critical points in the one-dimensional  soft-shoulder Hubbard model}

\author{M. Dalmonte}
    \affiliation{Institute for Theoretical Physics, University of Innsbruck, A-6020 Innsbruck, Austria}
    \affiliation{Institute for Quantum Optics and Quantum Information of the Austrian Academy of Sciences, A-6020 Innsbruck, Austria}
\author{W. Lechner}
    \affiliation{Institute for Theoretical Physics, University of Innsbruck, A-6020 Innsbruck, Austria}
    \affiliation{Institute for Quantum Optics and Quantum Information of the Austrian Academy of Sciences, A-6020 Innsbruck, Austria}
\author{Zi Cai}
    \affiliation{Institute for Quantum Optics and Quantum Information of the Austrian Academy of Sciences, A-6020 Innsbruck, Austria}
\author{M. Mattioli}
    \affiliation{Institute for Theoretical Physics, University of Innsbruck, A-6020 Innsbruck, Austria}
    \affiliation{Institute for Quantum Optics and Quantum Information of the Austrian Academy of Sciences, A-6020 Innsbruck, Austria}
\author{A. M. L\"auchli}
    \affiliation{Institute for Theoretical Physics, University of Innsbruck, A-6020 Innsbruck, Austria}
\author{G. Pupillo}        
\affiliation{ISIS (UMR 7006) and IPCMS (UMR 7504), University of Strasbourg and CNRS, Strasbourg, France}

\date{\today}

\begin{abstract}
We investigate the quantum phases of hard-core bosonic atoms in an extended Hubbard model where particles interact via soft-shoulder potentials in one dimension. Using a combination of field-theoretical methods and strong-coupling perturbation theory, we demonstrate that the low-energy phase can be a conformal cluster Luttinger liquid (CLL) phase with central charge $c=1$, where the microscopic degrees of freedom correspond to mesoscopic ensembles of particles. Using numerical density-matrix-renormalization-group methods, we demonstrate that the CLL phase, first predicted in [Phys.~Rev.~Lett.~111,~165302 (2013)], is separated from a conventional Tomonaga-Luttinger liquid by an exotic critical point with central charge $c=3/2$.  The latter is expression of an emergent conformal supersymmetry, which is not present in the original Hamiltonian. We discuss the observability of the CLL phase in realistic experimental settings with weakly-dressed Rydberg atoms confined to optical lattices. Using quantum Monte-Carlo simulations, we show that the typical features of CLLs are stable up to comparatively high temperatures. Using exact diagonalizations and quantum trajectory methods, we provide a protocol for adiabatic state preparation as well as quantitative estimates on the effects of particle losses.
\end{abstract}


\maketitle

\section{Introduction}

\input{intro_CLLGP.tex}

\input{clustersec.tex}

\input{Sec3b.tex}

\input{Sec4.tex}

%\section{Resilience of cluster features against temperature}

\input{cluster_liquid.tex}

\section{Adiabatic state preparation of a cluster Luttinger liquid state} \label{sec:adprep}

\input{adiabatic_1.tex}

\section{Conclusion and outlook}

\input{conclusions.tex}

\subsection*{Acknowledegments}

We acknowledge useful discussions with M. Fleischhauer, A. Gl\"atzle, C. Gross, F. Ortolani, M. Punk, and H. Weimer, and would like to thank
L. Huijse for correspondence on the nature of the critical point.
Numerical simulations for the adiabatic state preparation have been performed using QuTip libraries~\cite{qutip}.
Work in Innsbruck was supported in parts by the ERC Synergy Grant UQUAM, SIQS, EU Marie Curie ITN
COHERENCE, the SFB FoQuS (FWF Project No. F4006-N16), the ERA-NET CHIST-ERA (R-ION consortium), the Austrian Science Fund (FWF): P 25454-N27, and by the Austrian Ministry of Science BMWF as part of the UniInfrastrukturprogramm of the Focal Point Scientific Computing at the University of Innsbruck. Work in Strasbourg was supported by the ERC-St Grant ColdSIM (No. 307688), EOARD, UdS via IdEX, ANR via BLUESHIELD.

\end{document}

%% file: intro_CLLGP.tex
Frustration plays a fundamental role in our understanding of classical statistics mechanics.~\cite{magnet_book}
One remarkable example of the effects that frustration can induce on a many-body problem
is the self-assembly of conglomerate objects, or clusters. The emergence of such composite objects has been investigated in very different contexts, ranging from the physics and chemistry of colloidal particles and polymers \cite{LIKOS2001,MLADEK2006} to two- and three-dimensional bosonic systems of ultracold atoms and molecules~\cite{Cinti2013, Cinti2010, Henkel2010, Boninsegni2012}. There, the competition between superfluidity and clustering provides a mechanism to establish supersolidity. The latter corresponds to the simultaneous establishment of both diagonal and off-diagonal long-range order, a long-sought phenomenon in the context of quantum liquids~\cite{Boninsegni2012}.  

In the context of one-dimensional (1D) models, liquid phases are usually described by the Tomonaga-Luttinger liquid (TLL) universality class~\cite{haldane1981, GIAMARCHI2003, gogolin_book, cazalilla2004, cazalilla2011,testll}. The origin of such universality is rooted in the bosonization mapping, which allows for the reformulation of interacting bosonic, fermionic and spin models onto free bosonic theories. The latter can be exactly solved via path integrals as well as conformal field theories techniques~\cite{difrancesco_book}. TLLs are characterized by correlations decaying algebraically as a function of distance, implying quasi-long-range-order, where the precise values of the exponents depend crucially on the interactions in the microscopic Hamiltonian and on its symmetries, such as, e.g., the conservation of magnetization or total number of particles. TLLs are known to appear in a variety of realizations, including edges of topological phases~\cite{hall}, carbon nanotubes~\cite{carbon}, and cold gases of ions~\cite{Ion}, atoms~\cite{Atom} and molecules~\cite{Mol1, Mol2}. 

Recently, some of us have proposed that cluster formation can lead to a new class of quantum liquids, the so-called cluster Luttinger liquid (CLL),  presenting remarkably different features compared to TLL~\cite{PRL_MMMDWLGP}. One main point is that, in the CLL, the essential granularity in the liquid is given not by individual particles (as in regular TLL), but rather by {\it clusters} of particles. As a result, in the gapless CLL correlation functions still decay algebraically,  however, cluster features deriving from the underlying classical cluster structure remain evident. This results, for example, in a deformation of the Fermi surface similar to that taking place in Bose metals in ladder systems~\cite{Fisher2013, block2011, Mish2011}. This deformation leads to the appearance of features in structure factors and the momentum distribution that are not captured by TLLs - where Luttinger theorem always holds in the presence of a conserved U(1) symmetry associated with particle conservation~\footnote{In the non-particle conserving case, incommensurate features can appear, see, e.g., Ref.~\onlinecite{Sud2009}.}. 

It is the main aim of this work to shed further light onto the nature of the CLL state. In particular, we demonstrate that the CLL and the TLL are distinct phases with central charge $c=1$ and are separated by an exotic quantum phase transition that displays a quantum critical point with $c=3/2$. The latter is here associated to an emergent supersymmetry, for which we provide numerical evidence.  As a model, we focus on hard-core bosons trapped in a 1D geometry and interacting via a soft-shoulder potential. This kind of interaction can be treated exactly at the classical level, where it is shown to naturally lead to cluster formation. This provides an ideal starting point for the analysis of the role of quantum fluctuations based on this classical cluster phase. We further discuss in detail the possibility to observe these peculiar CLL states of matter in experiments with cold atomic gases, where soft-shoulder potentials can be engineered by coupling ground- and excited Rydberg states with laser light in the weak-dressing regime. There, both the range of the soft-shoulder interaction and its height can be efficiently controlled via external laser fields~\cite{Henkel2010,Pupillo2010,Pohl2014}. We conclude by presenting a scheme for the adiabatic preparation of the CLL in these systems and an analysis of their robustness under typical experimental conditions such as finite temperature.

The paper is organized as follows. In Sec.~\ref{sec:model_ham}, we present our model Hamiltonian, which is of the extended Hubbard type, and review its exact solution in the classical limit. In particular, we first provide a short summary of how soft-core potentials can be derived using cold Rydberg atoms~\cite{Henkel2010,Pupillo2010,Pohl2014} and then discuss the parameter regimes where mesoscopic clusters form (see Subsec.~\ref{subsec:cluex}). 

In Sec.~\ref{sec:cll}, we present a quantum mechanical analysis of the strong-coupling (Subsec.~\ref{subsec:strong}) and weak-coupling regimes (Subsec.~\ref{subsec:lowe}) of the Hamiltonian. In particular, using a combination of degenerate perturbation theory  from the classical cluster ground state and numerical calculations, in Subsec.~\ref{subsec:strong} we demonstrate that for large enough interaction strengths the effective low-energy dynamics is that of a gapless liquid of clusters. In Subsec.~\ref{subsec:lowe}, we complement this analysis by expanding the effective field theory approach presented in Ref.~\cite{PRL_MMMDWLGP} for the CLL phase, valid for weak interaction strengths. This latter approach retains all cluster constraints exactly and thus allows us to make predictions for the correlation functions of CLL (see Subsec.~\ref{subsec:corrf}). Both approaches reveal complementary aspects of the physics of cluster liquids and their differences with TLLs.

In Sec.~\ref{sec:phasd}, we analyze the model numerically using the density-matrix-renormalization group (DMRG) algorithm~\cite{White,Sch}. The latter is the state-of-the-art method to tackle 1D systems, and enables the quantitative investigation of the regime of intermediate interaction strengths, where analytical tools are not available. By a careful analysis of the scaling of observables such as the entanglement entropy (Subsec.~\ref{subsec:entr}) and the single-particle and cluster gaps (Subsec.~\ref{subsec:gaps}), we demonstrate that the TLL and the CLL are separated by a quantum phase transition described by an emergent $c=3/2$ conformal critical point~\cite{Dix88,difrancesco_book}. This is an interesting result, as the latter points toward the existence of an emergent supersymmetry, that is, a supersymmetry that is not explicitly present in the Hamiltonian. These types of critical points have a long history in the field of conformal field theories and have been recently discussed in the context of microscopic spin and fermionic models with many-particle constraints~\cite{Fen2003,Bauer}. In Subsec.~\ref{subsec:gaps}, a numerical analysis of low-lying excitations further demonstrates that the spectra of TLLs and CLLs are indeed very different. In fact, while in the TLL the single particle gap vanishes in the thermodynamic limit, in the CLL phase the gapless degrees of freedom correspond to a vanishing cluster gap. Interestingly, we find that the single particle gap opens linearly close to the transition point, which we discuss to further support the appearance of an Ising degree of freedom in the model, connected to the emergent supersymmetry. 

In Secs.~\ref{sec:temp} and~\ref{sec:adprep}  we discuss the stability of CLL phases in possible experiments with Rydberg-dressed atoms~\cite{revexp, exp, Gallagher, Santos2001, Saffman2010, Lahaye2009, Bloch2008, Baranov2012}. In particular, in Sec.~\ref{sec:temp}, we use quantum MonteCarlo (QMC) simulations to demonstrate that characteristic features of CLL are evident in correlation functions even at relatively high temperatures, signaling the stability of the CLL state against thermal fluctuations. In Sec.~VI we present a scheme for adiabatic state preparation of CLLs by numerically solving the master equation using a quantum jump approach~\cite{qj_zol,qj_dal,rev} for the interacting many-body system in the presence of dissipation. The main conclusion is that dissipation should not be detrimental when considering realistic time-scales for experiments with weakly-dressed Rydberg-atoms, which are usually affected by particle loss due to coupling to the Rydberg state. Finally, we draw our conclusions in Sec.~VII, and present the outlook of our work.

%% file: clustersec.tex
\section{Model Hamiltonian} \label{sec:model_ham}

\begin{figure}[t]
\begin{center}
\includegraphics[width = 0.98\columnwidth]{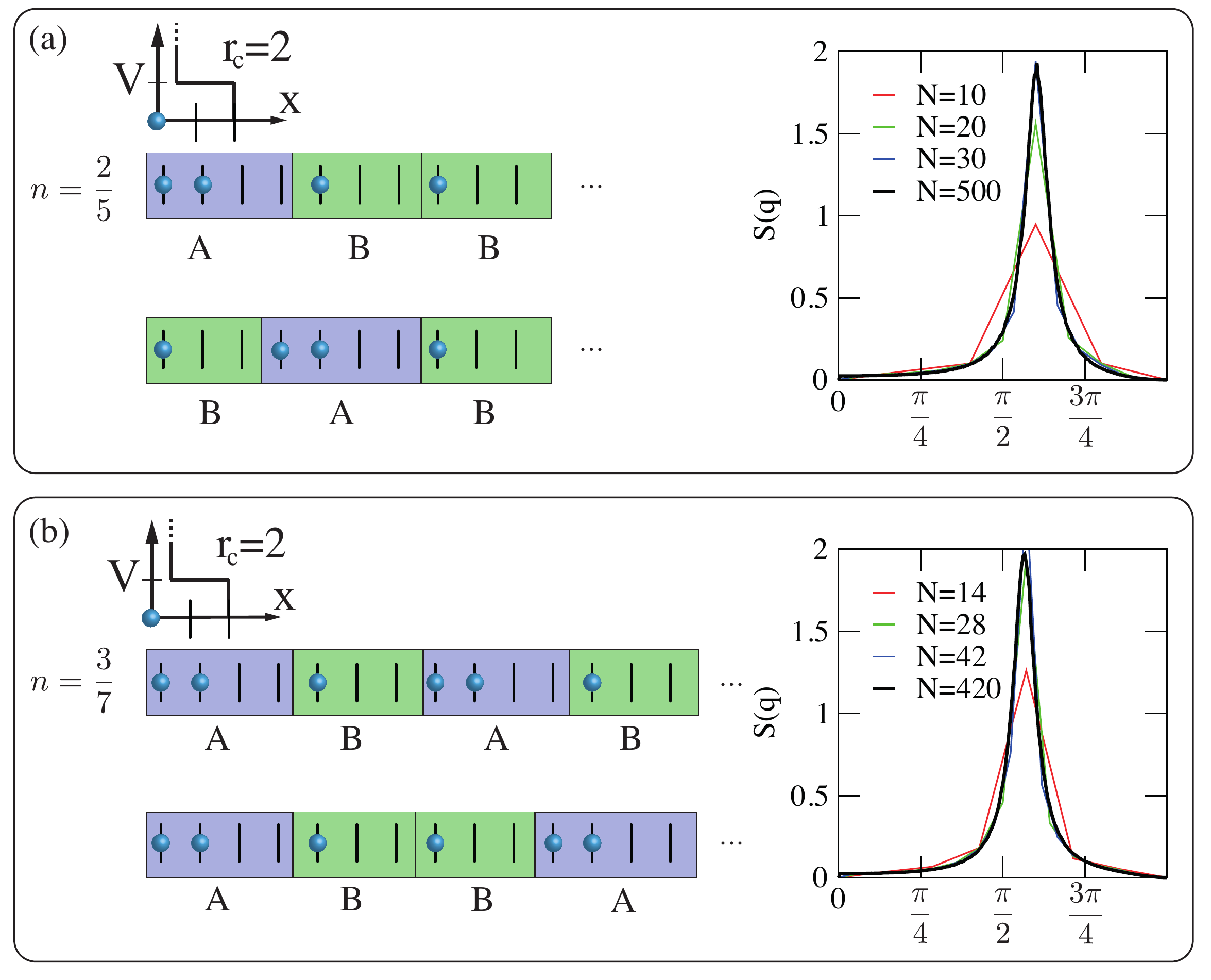}
\caption{Sketch of the degenerate ground state configuration for the Hamiltonian in Eq.~\eqref{equ:hamiltonian} in the classical regime ($t=0$) at density $n=2/5$ [panel (a)] and $n=3/7$ [panel (b)]. \textit{Insets}: The structure factor $S(q)$ versus the lattice quasi-momentum $q$ for various system sizes $N$. $S(q)$ is peaked at the $q$-vector characteristic of the formation of ground state cluster structures (see main text) and does not shift even for very small numbers of particles.  
\label{fig1}}  
\end{center}
\end{figure}

The model Hamiltonian we are interested in describes 
fermionic or hard-core bosonic particles in a 1D geometry and reads
\begin{eqnarray}
\label{equ:hamiltonian}
H&=&-t\sum_i (b^\dagger_ib_{i+1}+\mathrm{h.c.})+
V\sum_{i}\sum_{\ell=1}^{r_c}n_in_{i+\ell}.
\end{eqnarray}
Here, $b^{\dagger}_i (b_i)$ are bosonic/fermionic creation (annihilation)
operators at the site $i$, $n_i=b^{\dagger}_i b_i$, and $t$ is the tunneling rate. \\
%Moreover, we will consider density regimes which support cluster ground states in the classical and quantum limit. \\

Interactions between particles have a soft-core profile, with depth $V$ and radius $r_c$. 
These particular features of the interaction potential can be experimentally realised with Rydberg-dressed cold gases. In the weak-dressing regime $\Omega \ll |\Delta|$, atoms in their ground states are off-resonantly coupled with Rabi frequency $\Omega$ and red detuning $\Delta<0$ to an high-lying Rydberg state. The resulting effective potential as a function of the relative distance $x$ between pair of atoms reads~\cite{Henkel2010}

\begin{equation}
V(x) = \frac{\Omega^4}{8\Delta^3}\frac{r_c^6}{r_c^6 + x^6}
\label{equ:Vx}
\end{equation}  
where $r_c = [C_6/(2|\Delta|)]^{1/6}$ is the Condon radius and $C_6$ is the van-der-Waals coefficient of the addressed Rydberg state. At large distances $x \gg r_c$, $V(x)$ reduces to the usual repulsive van-der-Waals interaction between Rydberg atoms $\propto x^{-6}$, suppressed by a factor $[\Omega/(2\Delta)]^4$, since only a small fraction $[\Omega/(2\Delta)]^2$ of the Rydberg state is admixed into the original ground state. However, for $x<r_c$, a double Rydberg excitation is prevented by the dipole blockade and $V(x)$ saturates to a universal constant value $\Omega^4/8\Delta^3$. 

\subsection{Cluster exchange model} \label{subsec:cluex}

Before presenting results for the quantum phases of Hamiltonian Eq.~(\ref{equ:hamiltonian}), we first summarize results for the classical ground state. Particles with soft-shoulder interactions and their cluster-phases have been extensively studied in the classical regime in the context of soft matter physics \cite{LIKOS2001,MLADEK2006}. In one dimensional chains, we study the classical ground state from applying the \textit{cluster exchange model} introduced in Ref.~\onlinecite{PRL_MMMDWLGP}. The two relevant length scales in this regime are the cut-off radius $r_c$ and the average distance between particles $r^\star=1/n$, with $n=N/L$ the particle density. Here, $N$ and $L$ are the number of particles and of lattice sites respectively. This leads to three possible regimes in the classical limit: (i) liquid $r^\star > r_c$, (ii) crystal $r^\star = r_c$ and (iii) cluster liquid phase $r^\star < r_c$. In this work we will consider $r_c = 2$ and densities $n=2/5$ and $n=3/7$ which both correspond to a cluster phase. 

The cluster exchange model is illustrated in Fig.~\ref{fig1}: For, e.g., $r_c=2$ and $n=2/5$ the state with lowest energy consists of single particles and clusters of size $2$ [depicted in Fig.~\ref{fig1}(a)] with a total energy of $E_0 = V n$. However, the ground state is not a unique configuration but there are exponentially many degenerate configurations with the same energy. These configurations can be represented by dividing the system in blocks made of particles and holes, where the number of holes is the same in each block. Here, e.g., blocks labeled $A$ consist of two particles and two holes with a total length of $4$, whereas  blocks labeled $B$ consist of one particle and two holes with length $3$. In this block model, each density of particles $n$ corresponds to a ratio of number of blocks $A$ ($B$), $N_A (N_B)$. For example, $n=2/5$ corresponds to a ratio $N_A/N_B=1/2$ and $n=3/7$ to $N_A/N_B=1$, respectively. The ground state then consists of all permutations of blocks A and B (e.g. for $n=3/7$ typical configurations are [AABAAB...], [ABAAAB...], $...$)\cite{PRL_MMMDWLGP}. The associated ground state degeneracy, assuming open boundary conditions, is $d = M!/[(M/3)! (2M/3)!]$ for $n=3/7$ and $d=M!/[(M/2)!]^2$ for $n=2/5$, respectively, where $M=(L-N)/r_c$ is the total number of clusters in the system.

%% file: Sec3b.tex
\section{The cluster Luttinger liquid phase} \label{sec:cll}

The complex structure of the classical ground states results in the emergence of an exotic quantum liquid once quantum fluctuations are introduced. In this Section, we will first present a strong coupling approach in the $V\gg t$ limit, and then a modified bosonization treatment which embodies the cluster constraints derived in the classical limit. The combination of the two approaches allows us to gain a qualitative analytical understanding of the full phase diagram of Eq.~\eqref{equ:hamiltonian}, which we quantitatively investigate in Sec.~IV below.

\subsection{Strong-coupling approach to cluster manifolds: the XX model} \label{subsec:strong}

Once the degenerate manifold of clustered ground states has been identified, it is possible to derive an effective Hamiltonian describing the dynamics in the limit $t\ll V$. In order to do so, we define as $\mathcal{P}$ the projector operator on the classical ground state manifold, and apply conventional second order perturbation theory (odd order corrections are not present for the case $r_c=2$)
\begin{equation}
H_{\textrm{eff.}} \simeq H_0 + \mathcal{P}H_t \mathcal{Q}\frac{1}{\epsilon - H_0} \mathcal{Q}H_t \mathcal{P} + \mathcal{O}(t^4/V^3),
\end{equation}
where $\mathcal{Q}= \mathbb{1} - \mathcal{P}$, and $\epsilon$ is the classical ground state energy within the cluster manifold. First, we define effective spin-1/2 operators $\tilde{S}_j$ as follows: Ordering the cluster configuration with an index $j\in[1,M]$, we associate to the position of each $A$-type cluster a spin-up, and for each $B$-type, a spin-down. This is a one-to-one mapping of the Hilbert space defined by $\mathcal{P}$ to the Hilbert space of a spin-1/2 chain with $M$ sites. The effective Hamiltonian can be then cast in a compact form as a spin chain. For the $r_c=2$ case, diagonal contributions of the type $\tilde{S}^z_i\tilde{S}^z_{i+1}$ do not contribute at lowest order and we get
\begin{equation}\label{2order}
H_{\textrm{eff.}} \simeq H_0 -\frac{t^2}{V}\sum_{j = 1}^{M}[ (\tilde{S}^+_j\tilde{S}^-_{j+1} + h.c.) +2].
\end{equation}
The strong coupling limit can then be mapped to a system of hard-core bosons (spin-1/2) hopping in an artificial lattice created by the underlying cluster structure. This confirms that the CLL, which is adiabatically connected to the strong coupling limit, is indeed described at low-energies by a $c=1$ conformal field theory (CFT), which is interpreted as a Luttinger liquid of composite cluster particles~\cite{PRL_MMMDWLGP}.

\begin{figure}[t]
\begin{center}
\includegraphics[width = 0.98\columnwidth]{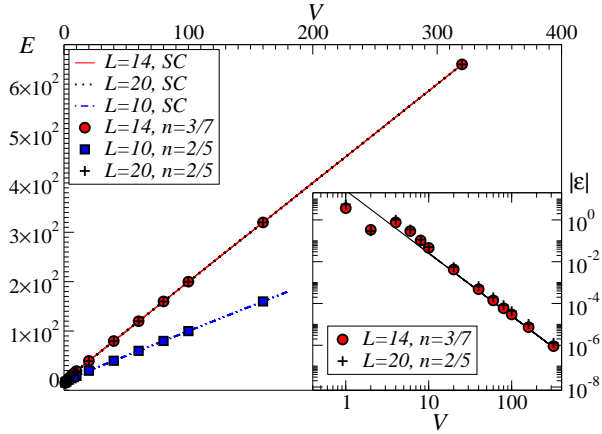}
\caption{{\it Main Panel}: Energy of the ground state (GS) as a function of $V/t$ in the perturbative limit $t\ll V$ (here $t=1$). Symbols denote the exact GS energies obtained via exact diagonalization of Eq.~\eqref{equ:hamiltonian}, while lines denote perturbative estimates for different cluster configurations and system sizes. 
{\it Inset:} Absolute value of the energy difference $\epsilon$ between the exact and the perturbative estimate. The energies are quite accurate down to low values of $V/t \lesssim 10 $, where the strong coupling expansion around the cluster manifold first starts to break down. The thin, black line is a guide for the eye indicating a power law decay $t^4/V^3$. The latter demonstrates that higher order terms are irrelevant at large couplings, but play a role around $V/t \simeq 10$. The small kink around $V/t \simeq 2$ is due to the fact that $\epsilon$ changes sign in this region.
\label{EDcomp}}  
\end{center}
\end{figure}

In order to benchmark the validity of the strong coupling expansion and to check to which extent corrections to Eq.~(\ref{2order}) are quantitatively relevant, we compare in Fig.~\ref{EDcomp} the energy of a system of $L=14$ and $10, 20$ sites for $n=3/7$ and $2/5$, respectively, and $r_c=2$, using exact diagonalizations of the full Hamiltonian as well as perturbative estimates. We find a very good agreement between the exact and perturbative results down to relatively small values of $V/t\simeq 5$. This indicates that all relevant quantum dynamics is well described within our strong-coupling model detailed above, where clusters play the role of the mesoscopic degrees of freedom. At smaller interaction values, higher order terms are non-negligible, and moreover the cluster assumption breaks down, as demonstrated by the poor agreement between the exact and perturbative results below $V/t\simeq5$ [see inset in Fig.~\ref{EDcomp}].

\subsection{Beyond perturbation theory: Low-energy field theory of cluster Luttinger liquids} \label{subsec:lowe}

While perturbation theory provides an understanding of the large $V$ limit, an analytical picture at intermediate couplings is hindered by the complex structure of the interactions. In Ref.~\onlinecite{PRL_MMMDWLGP}, we discussed how cluster-type constraints can be generically applied to Haldane's bosonization prescription to derive a modified mapping between the original microscopics fields and continuous bosonic variables. The main feature of this treatment is that it captures the correct behavior of correlation functions and the deformation of the Bose surface, which is instead not accessible by direct bosonization of Eq.~\eqref{equ:hamiltonian}.  Here, we provide a detailed derivation of the mapping and a discussion of its consequences on various observables. Analytical predictions will be compared with exact numerical simulations in the next sections.\\

We start from the second step of Haldane's construction~\cite{haldane1981,GIAMARCHI2003}, by considering a more
complicated shape of the initial particle distribution that can reflect the underlying cluster 
structure. After taking the continuum limit, this can be cast as the following constraint on the density distribution for the density operator $\rho$
\begin{eqnarray}
\rho(x)&=&\sum_{n=1}^{N}\delta(x-x_n)\doteq \sum_{m=1}^M\delta(x-x_m)+
\sum_{\ell\in cl}\delta(x-x_{\ell})=\nonumber\\
&=&\sum_{m=1}^Mf(x_m)\delta(x-x_m).
\end{eqnarray}
Here, $\sum_{x_m}f(x_m)=N$, and the function 
$f(x_m)\in \{1,2\}$ describes both one- and two-particle cluster structures. 
In the center equality, $\sum_{cl}$ represents the sum over 
all two-particle clusters.
Formally, the summation $\sum_{cl}$ constraints the values of $x_\ell$ such 
that there exists an $m$ such that $x_\ell=x_m\;\forall \ell$. The key point here is that the sum is split into two parts: the first one describes the fact that there are some clusters where at least one particle resides, while the second sum takes into account the possibility of having clusters formed by two particles. We now introduce a new field, $\varphi_{cl}(x)$, which accounts for quantum fluctuations of the cluster density. The field satisfies:
\begin{equation}
\varphi_{cl}(x)=\varphi_{cl}(x+L)+M \pi, \quad \varphi_{cl}(x_m)=\pi m %\quad \frac{N}{M}\nabla\varphi_{cl}(x)=\pi (n+\Pi(x)),
\end{equation}
similarly to the standard density fluctuations in the Haldane scenario.
Now, we can apply the standard representation 
of the delta function
\begin{equation}\label{eqdelta}
\delta[g(x)]=\sum_{\textrm{zeros of } g}\frac{1}{|\nabla g(x_j)|}\delta(x-x_j)
\end{equation}
which, when combined with the previous ansatz, gives
\begin{eqnarray}
\rho(x)&=&\sum_{m=1}^Mf(x_m)\delta(x-x_m)=\nonumber\\
&=& \nabla\varphi_{cl}(x)\sum_{m=1}^{M}f(x_m)\delta[\varphi_{cl}(x)-\pi m] .
\end{eqnarray}
The initial density formula is recovered by applying to the right-hand-side of the latter formula 
the delta function transformation of Eq.~\eqref{eqdelta}. 
The next passage is then to Fourier transform the previous formula.
This is rather different with respect to the Haldane scenario; in the 
latter, one has to Fourier transform a standard Dirac comb. Here, 
the Dirac comb is in fact \emph{weighted} by the cluster configuration
$f(x_m)$, which affects its Fourier components. Let us define
the functional:
\begin{eqnarray}\label{chi}
\chi[\varphi_{cl}(x)]&=&\sum_{m=1}^{M}f(x_m)\delta[\varphi_{cl}(x)-\pi m] =  \\
& = &\sum_{m=1}^{M}\delta[\varphi_{cl}(x)-\pi m]  
 +  \sum_{\ell\in cl} \delta[\varphi_{cl}(x)-\pi \ell]\nonumber ,
\end{eqnarray} 
where the last sum is again performed on two-particle clusters. The first 
part can be Fourier transformed by considering the Poisson summation
formula
\begin{equation}
\sum_{n=-\infty}^{\infty}\delta(x-nK)=\sum_{k=-\infty}^{\infty}\frac{1}{K}e^{-i\pi kx/K}
\end{equation}
thus leading to 
\begin{equation}
\lim_{M\rightarrow \infty}\sum_{m=1}^{M}\delta[\varphi_{cl}(x)-\pi m] = \sum_{k=-\infty}^\infty \frac{1}{\pi}e^{-i k\varphi_{cl}(x)}.
\end{equation} 
The last important point is to re-absorb the last term in Eq.~\eqref{chi}.
By Fourier expanding all of its components, one gets
renormalized $c_k$ coefficients in the previous expression - not
affecting its functional form. As such, we neglect these effects in the 
following, noticing that they are nevertheless expected to be very small
when the number of two-particle clusters is small, that is, $M\simeq N$.
We will then re-express the particle density as a function of the
cluster operators as follows 
\begin{equation}
\rho(x)=\frac{N}{M}\nabla\varphi_{cl}(x)\left\{\sum_{k=-\infty}^\infty \frac{a_k}{\pi}e^{-i k\varphi_{cl}(x)}, \right\}
\end{equation}
where the numerical pre-factor has been introduced to compensate 
for the delta function renormalized coefficients. We now proceed by rescaling the fields as follows:
\begin{equation}
\varphi_{cl}(x)=-2\varphi_{cl}(x)'+2\pi n \sigma x,
\end{equation}
where the new field $\varphi_{cl}(x)'$ represents fluctuations on top of the 
perfect cluster-crystalline solution of the problem, and $\sigma=M/N$.
In this way we obtain the final form of the mapping between the microscopic continuum density and the cluster fields
\begin{eqnarray}\label{rho_cl}
\rho(x)&=&\frac{N}{M}\nabla\varphi_{cl}(x)\left\{\sum_{k=-\infty}^\infty \frac{1}{\pi}e^{-i k\varphi_{cl}(x)} \right\}=\\
&=&\left[ n- \frac{\sigma}{\pi} \nabla \varphi_{cl}(x)' \right]
\left\{\sum_{k=-\infty}^\infty a_ke^{2i k[\varphi_{cl}(x)'-\pi n\sigma x]} \right\}.\nonumber
\end{eqnarray}
The single particle operators can also be expressed by applying 
the same procedure of the Haldane formalism. In the bosonic case, 
we get:
\begin{eqnarray}\label{psiB}
\psi^B(x)\simeq \sqrt{n- \frac{\sigma}{\pi} \nabla \varphi_{cl}(x)' }
e^{-i\beta\vartheta_{cl}(x)'}\times\nonumber\\
\times \left\{\sum_{k=-\infty}^\infty \alpha_ke^{i2 k[\varphi_{cl}(x)'-\pi n\sigma x]} \right\},
\end{eqnarray}
where the operator $\vartheta_{cl}(x)'$ is the conjugate of $\varphi_{cl}'$,
satisfying
\begin{equation}\label{comm1}
\left[\frac{1}{\pi}\nabla\varphi_{cl}(x)',\vartheta_{cl}(y)'\right]=-i\delta(x-y).
\end{equation}
This can be verified by considering the role of the factor $\beta$;
approximating the density operators with its non-oscillating part, 
we need the following commutation to hold
\begin{equation}
\left[\frac{\sigma}{\pi}\nabla\varphi_{cl}(x)',e^{-i\beta\vartheta_{cl}(y)'}\right]=-\delta(x-y)e^{-i\beta\vartheta_{cl}(y)},
\end{equation}
which is satisfied if $\beta=\sigma^{-1}$ and Eq.~\eqref{comm1} holds.
\begin{figure}[t]
\begin{center}
\includegraphics[width = 0.88\columnwidth]{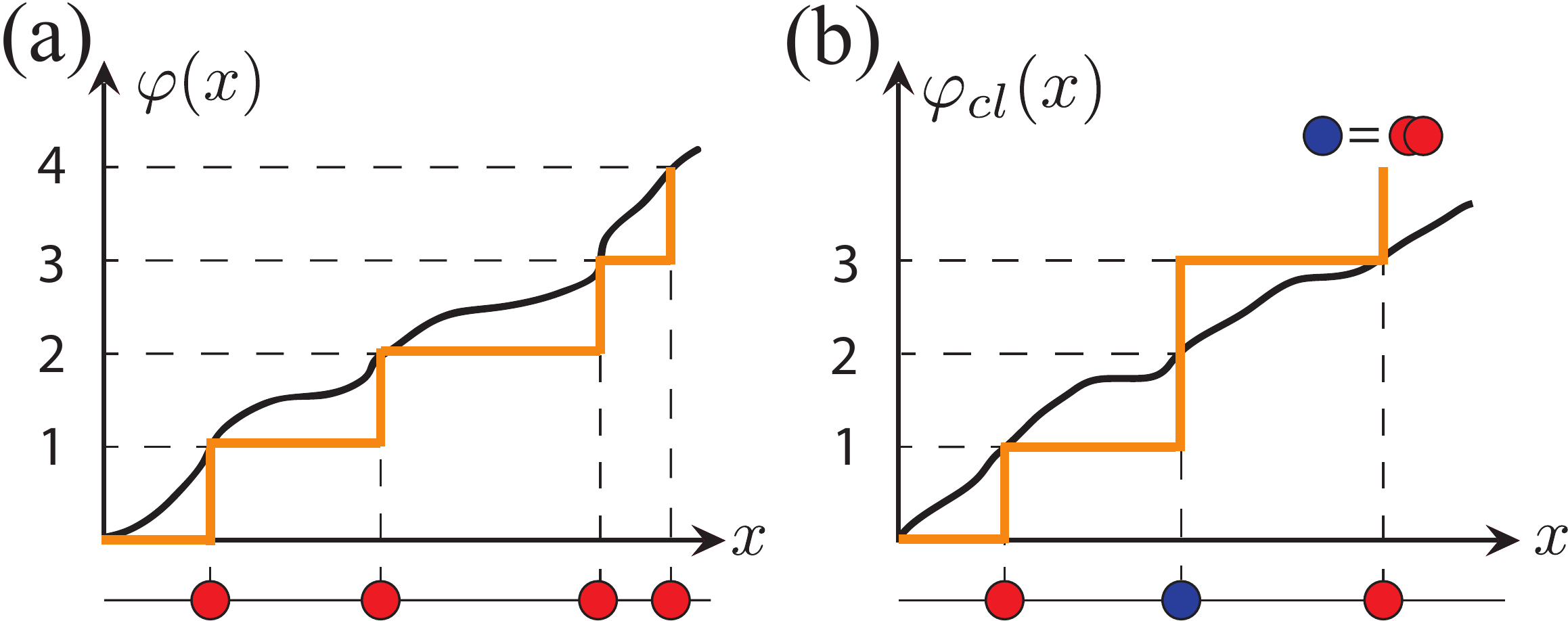}
\caption{Configurations allowed for the field $\varphi(x)$. Panel (a): In the Luttinger
liquid scenario, $\varphi(x)$ can describe all possible particle configurations and takes
integer values at the position of each particle (red circles). The integrated particle density (thin orange line) follows the profile of $\varphi$. Panel (b): In the cluster Luttinger liquid scenario, $\varphi_{cl}$ is constrained by the cluster structure, which assumes that a finite
number of particles are \emph{packed} into small clusters (blue circles). The integrated particle density (thick orange line) does not follow the behavior of $\varphi_{cl}$,  but jumps whenever a two-particle cluster is encountered. 
\label{TLL_panelv2}}  
\end{center}
\end{figure}
The effective low-energy Hamiltonian obtained by introducing the cluster fields in the microscopic Hamiltonian is thus a compactified boson theory
\begin{equation}
\mathcal{H} = \frac{v_{cl}}{2}\int dx \left[ (\partial_x\varphi_{cl})^2/K_{cl} + K_{cl},(\partial_x\vartheta_{cl})^2\right]
\end{equation}
which is gapless and conformal, with central charge $c=1$. All correlation functions of the microscopic operators can be evaluated using conventional techniques - see Ref.~\cite{difrancesco_book} for a review. The theory is thus very similar to a TLL, with the exception that the mapping between microscopic and low-energy degrees of freedom presents remarkable differences. In the next subsection, we will illustrate the effects on correlations, while in Sec.~\ref{sec:phasd}, we will show numerically that the interpretation of the low-energy excitations differs considerably in a CLL with respect to the TLL.

\subsection{Correlation functions and structure factors in the cluster Luttinger liquid phase} \label{subsec:corrf}

The mapping between the original operators and the emergent cluster fields in~\eqref{rho_cl} and~\eqref{psiB} allows us to make predictions on the scaling of both Green's functions and density-density correlations. As an example, we focus here on the latter (the effects on the former have been discussed in Ref.~\cite{PRL_MMMDWLGP}), which after bosonization take the form
\begin{equation}
\langle\rho(x)\rho(0)\rangle\simeq n^2+\frac{\alpha_1}{x^2}+
\frac{\alpha_2\cos(2\pi n\sigma x)}{x^{\gamma_1}}+... \;,
\end{equation}
with $\alpha_1$, $\alpha_1$ and $\gamma_1$ non-universal coefficients. This latter expression displays a radically different spatial modulation with respect to the 
standard Luttinger liquid scenario. In particular, assuming periodic boundary conditions (PBCs) peaks in the static structure factor
\begin{equation}\label{Sq}
S(q)=\frac{1}{L}\sum_{\ell, j}e^{i(\ell-j)q}\left[\langle n_\ell n_j\rangle -n^2\right]
\end{equation}
are now {\it not} displayed at momenta associated with the particle density, as in TLLs. In particular, the lowest momentum peak is located at 
\begin{equation} \label{char_wv}
k_1 =\frac{2\pi(1-n)}{r_c}
\end{equation}
which for $r_c =2$ , gives $k_1 = \pi (1-n)$. This is exactly the corresponding momentum peak of the classical cluster configuration, which is unrelated to the density of individual particles as in TLLs. We note that other signatures of departure from the TLL picture can also be found using level spectroscopy techniques~\cite{Zhuravlev2000}.

In the next two sections, we provide numerical evidence for the existence of a CLL at intermediate couplings in Eq.~\eqref{equ:hamiltonian}, by monitoring the entanglement entropy, the spectral properties, and correlation functions.

%% file: Sec4.tex
\section{Phase diagram} \label{sec:phasd}

In this section, we investigate the phase transition from the TLL to the CLL phase by using numerical simulations based on the DMRG algorithm~\cite{White,Sch}. As boundary effects are expected to be prominent in the cluster phase, we perform simulations using PBCs for systems up to $L=70$ sites, keeping up to 1400 states per block and with up to 10 finite-size sweeps. The typical truncation error in the final sweep is of the order $2\cdot 10^{-6}$ for $L<60$ and $<  10^{-5}$ for $L=70$. Observables such as the local density $n_j$ deviate from the mean value $n$ at most as $n_j-n< 3\cdot10^{-5}$.

In order to illustrate the generality of our findings, we investigate two relevant scenarios for the case $r_c=2$, namely, different densities $n=3/7$ and $n=4/10$, which in the classical limit lead to different cluster densities as discussed in the previous sections (density regimes where dominant Umklapp terms appear have been investigated in Ref.~\cite{Schmitteckert2004}). In order to keep commensurability with the cluster structure, we considered sizes of $(14, 28, 42, 56, 70)$ and $(20, 30, 40, 50, 60, 70)$ for the two cases, respectively.

\subsection{Entanglement entropies} \label{subsec:entr}

In order to locate the transition point between the TLL and the CLL, as a first observable we monitor how the ground state entanglement properties change as a function of $V/t$. We consider the bipartite von Neumann entanglement entropy
\begin{equation}
S_A = - \textrm{Tr}\rho_A\log\rho_A,
\end{equation}
where $\rho_A$ is the reduced density matrix of the sub-system $A$ with respect to the rest of the chain. As both the TLL and the CLL are described at low-energies by a conformal field theory, the entanglement entropy fulfills the following scaling~\cite{Wil,calabrese}
\begin{equation}\label{eqS2}
S_L(l) = \frac{c}{3} \ln \left[\frac{L}{\pi}\sin(\pi l /L)\right] + C + \mathcal{O}(1/l^\alpha),
\end{equation}
where $L $ is the system size, $l$ is the block length, $C$ is a non-universal constant, and $c$ is the central charge of the theory. Corrections of order $1/l^\alpha$, if any, are expected to exhibit no oscillations for the von Neumann entropy under PBCs. 

\begin{figure}[t]
\begin{center}
\includegraphics[width = 0.49\columnwidth]{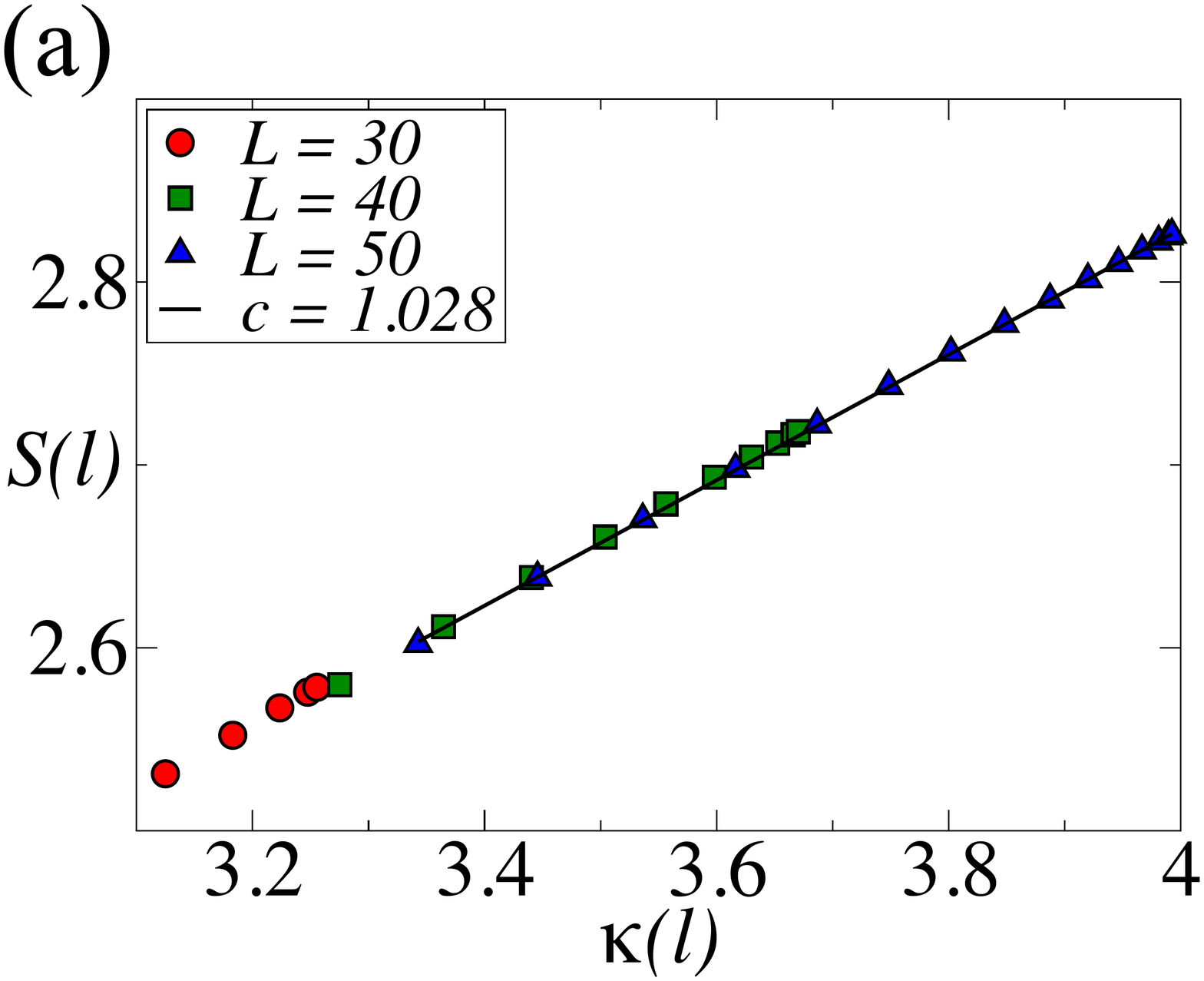}
\includegraphics[width = 0.49\columnwidth]{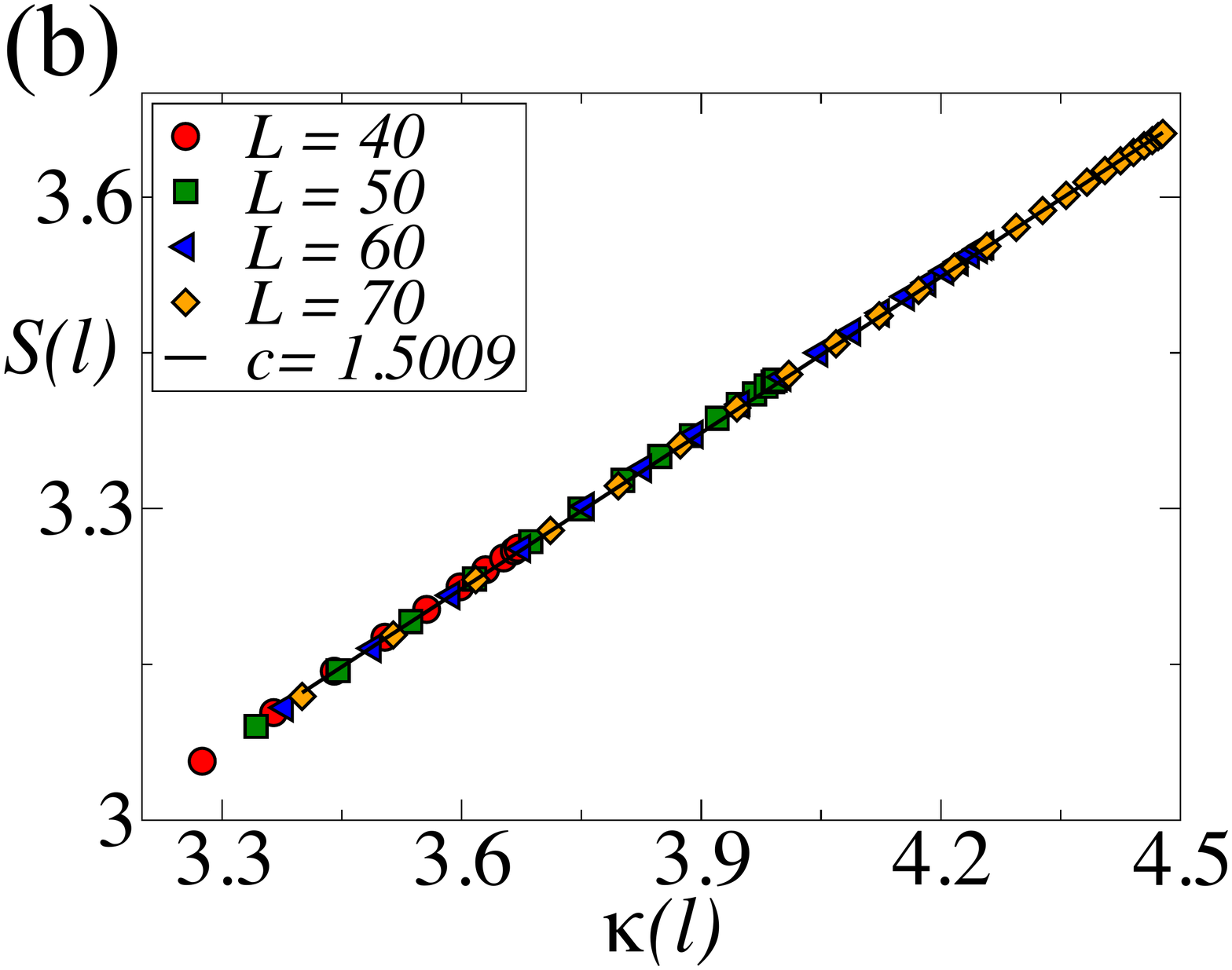}
\includegraphics[width = 0.49\columnwidth]{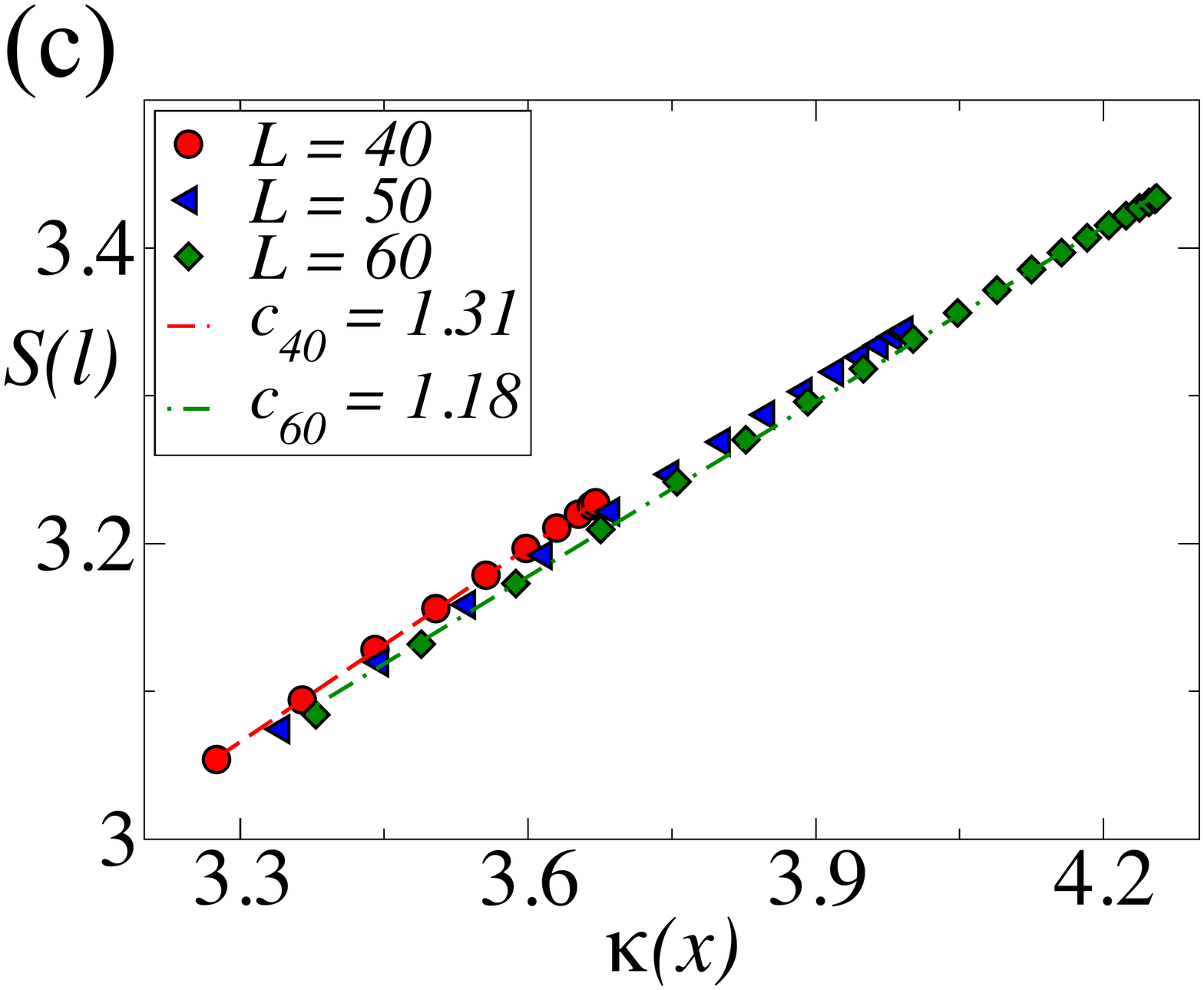}
\includegraphics[width = 0.49\columnwidth]{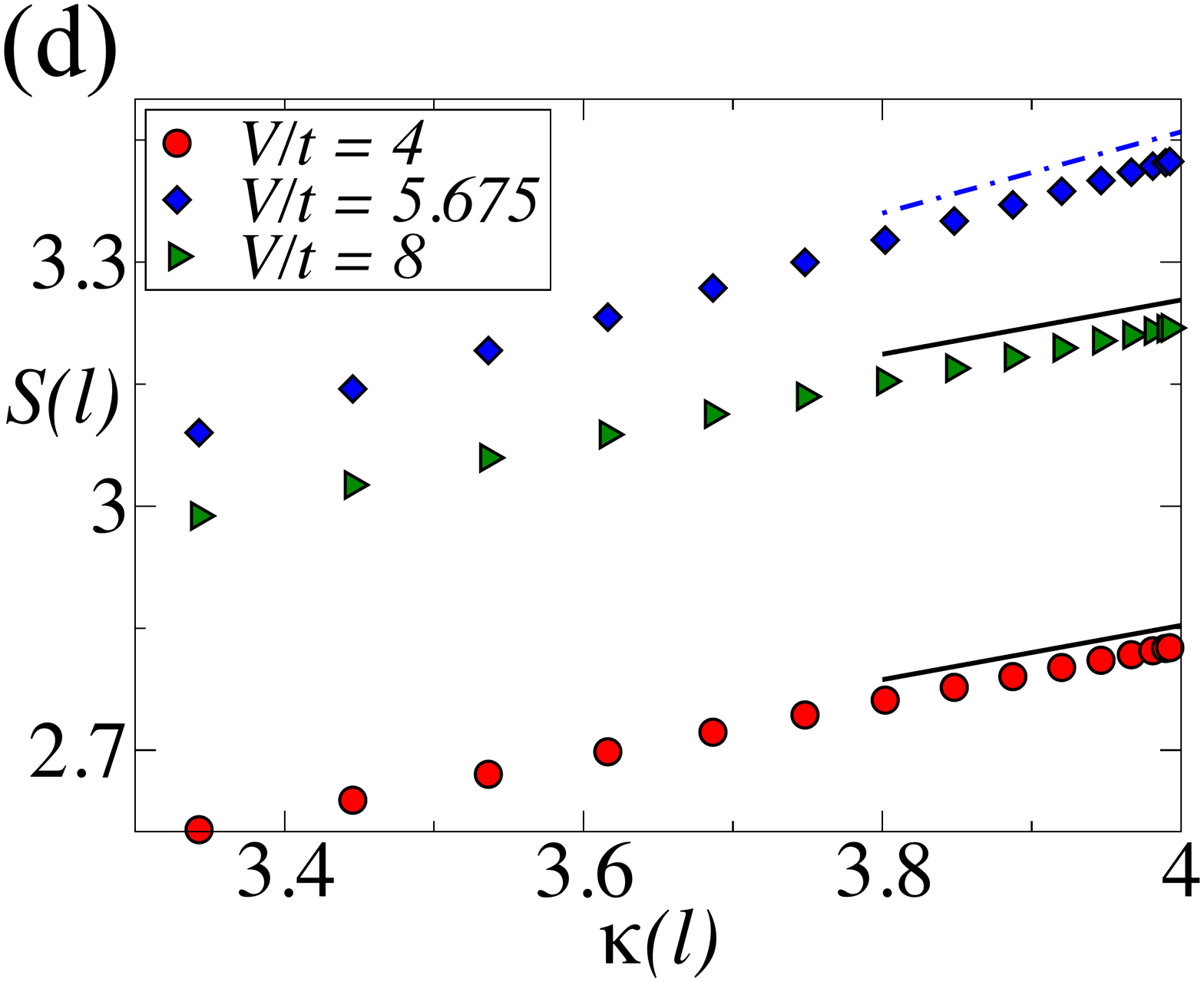}
\caption{Entanglement entropy scaling as a function of the block length for $n=4/10$. Panels (a)-(c): different interaction strengths $V/t = 4.0, 5.675, 6.5$. Data for different sizes are indicated by symbols, while lines are linear fits. Deep in the TLL phase, the finite-size central charge quickly approaches $1$. At the transition point, the value is $1.5$. Deep in the CLL phase, finite-size-effects are strong - as evidenced by the bending of the entropy for different system sizes. 
Panel (d): $L=50$, different interaction strengths. The black continuous and blue dot-dashed lines are guides for the eye at $c = 1$ and $3/2$, respectively. 
\label{fig_S}}  
\end{center}
\end{figure}

\begin{figure}[t]
\begin{center}
\includegraphics[width = 0.49\columnwidth]{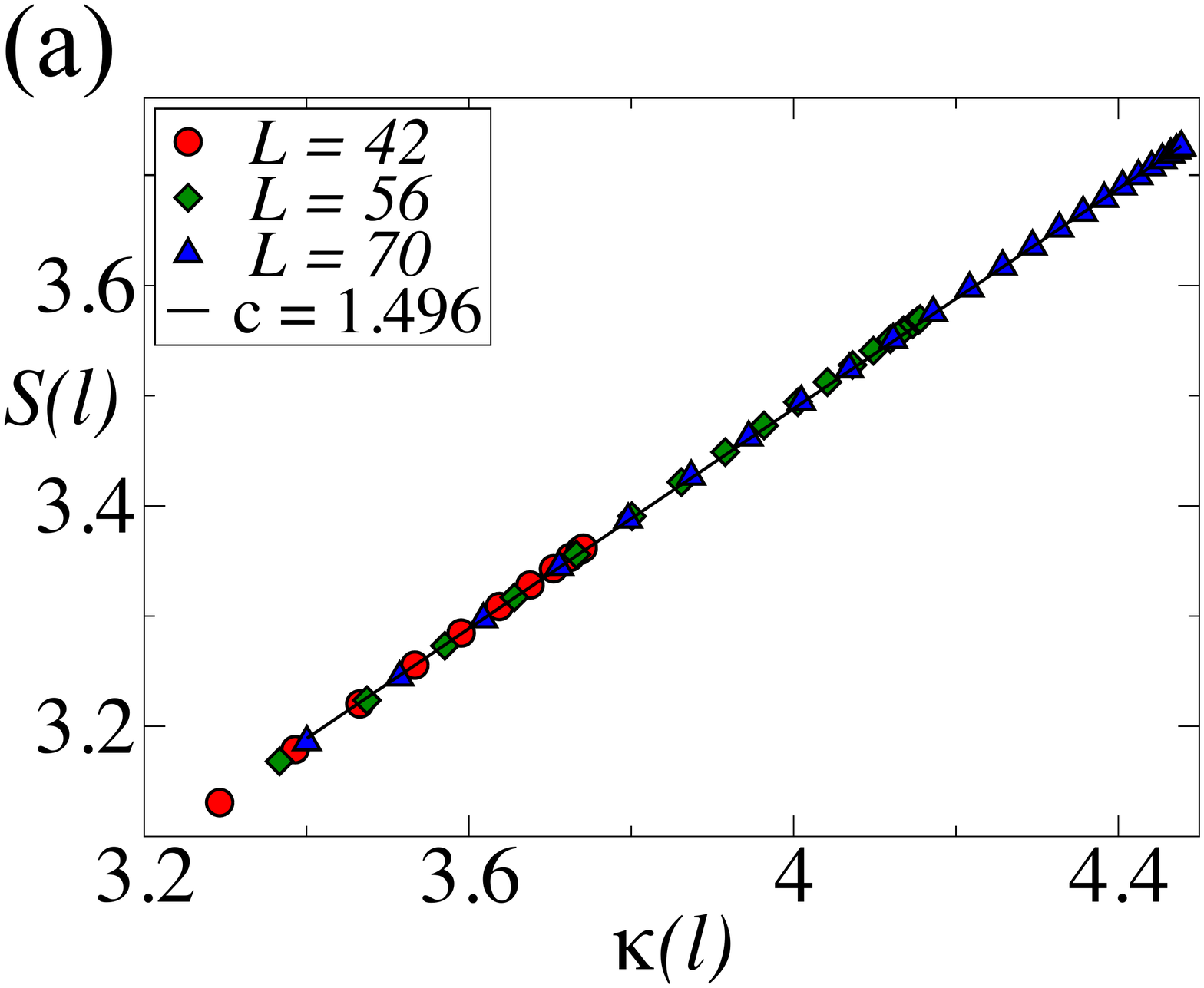}
\includegraphics[width = 0.49\columnwidth]{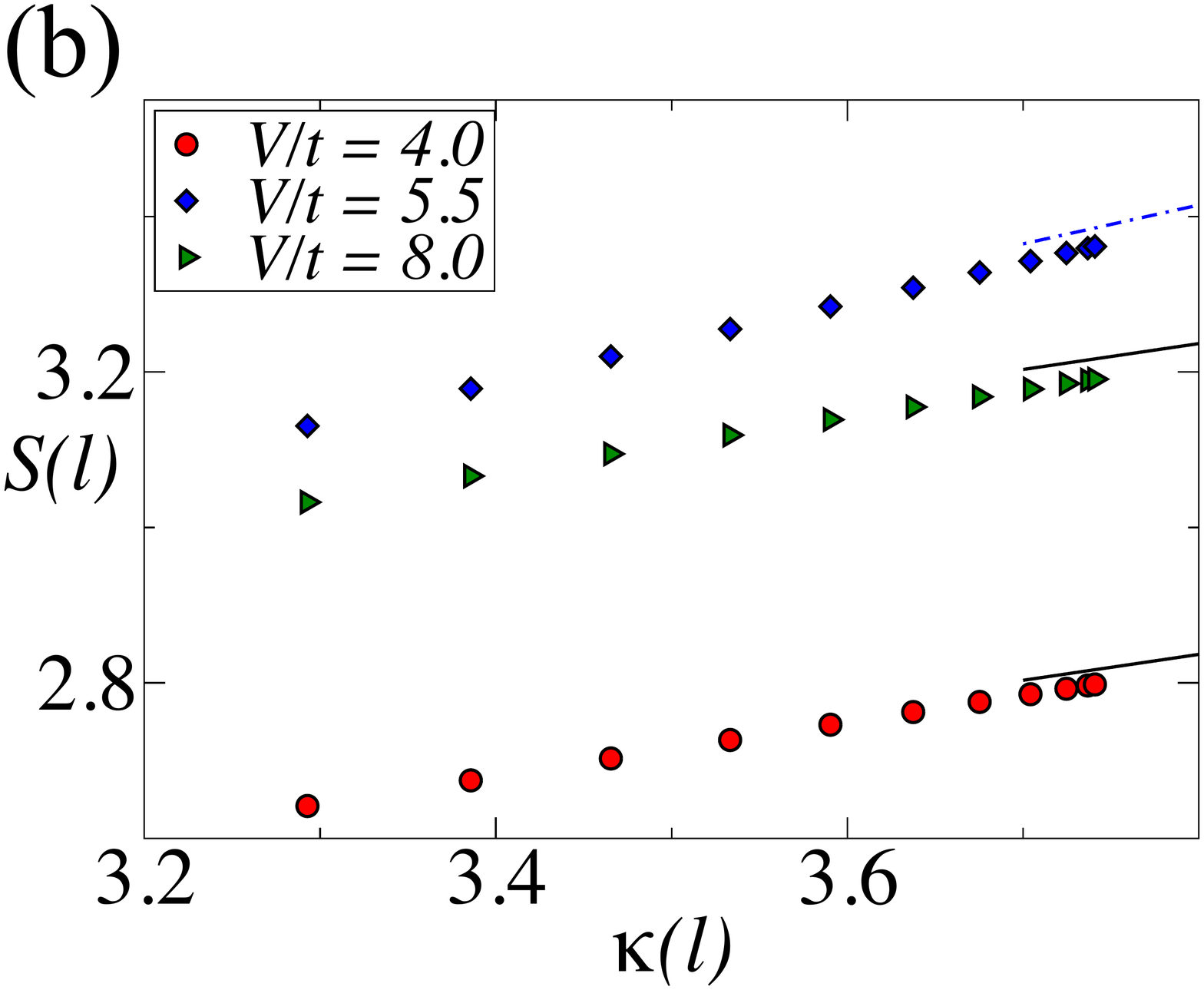}
\caption{Entanglement entropy scaling as a function of the block length for $n=3/7$. Panel (a): entropies at the transition point $V/t = 5.5$. Data for different sizes are indicated by symbols, while the black line is a linear fit in good agreement with $c=3/2$. Panel (b): $L=50$, different interaction strengths. The black continuous and blue dot-dashed lines are guides for the eye at $c = 1$ and $3/2$, respectively. 
\label{fig_S2}}  
\end{center}
\end{figure}

In order to extract the central charge of the system from finite-size simulations, we proceed in two-steps. First, we evaluate the finite-size value of the central charge, $c(L)$, by fitting the entanglement entropy at a fixed size with the following function
\begin{equation}
S_L(l) = \frac{c(L)}{3} \kappa(l) + a_0,\quad  \kappa(l) = \log\left[\frac{L}{\pi}\sin(\pi l /L)\right],
\end{equation}
where $\kappa(l)$ is the logarithm of the cord length. We keep $l=7$ as the minimum block size considered in order to avoid possible corrections due to the breakdown of Eq.~\eqref{eqS2} for blocks that are too small. Typical results are presented in Figs.~\ref{fig_S} and~\ref{fig_S2}. We find that a linear fit works quite well for all system sizes and in the entire parameter regimes we investigated. The finite-size central charge can be extracted with an accuracy of order $1\%$\footnote{The error is estimated by performing fits including additional $1/l^\alpha$ corrections, and excluding the smaller blocks from the fitted data.}. 

\begin{figure}[t]
\begin{center}
\includegraphics[width = 0.85\columnwidth]{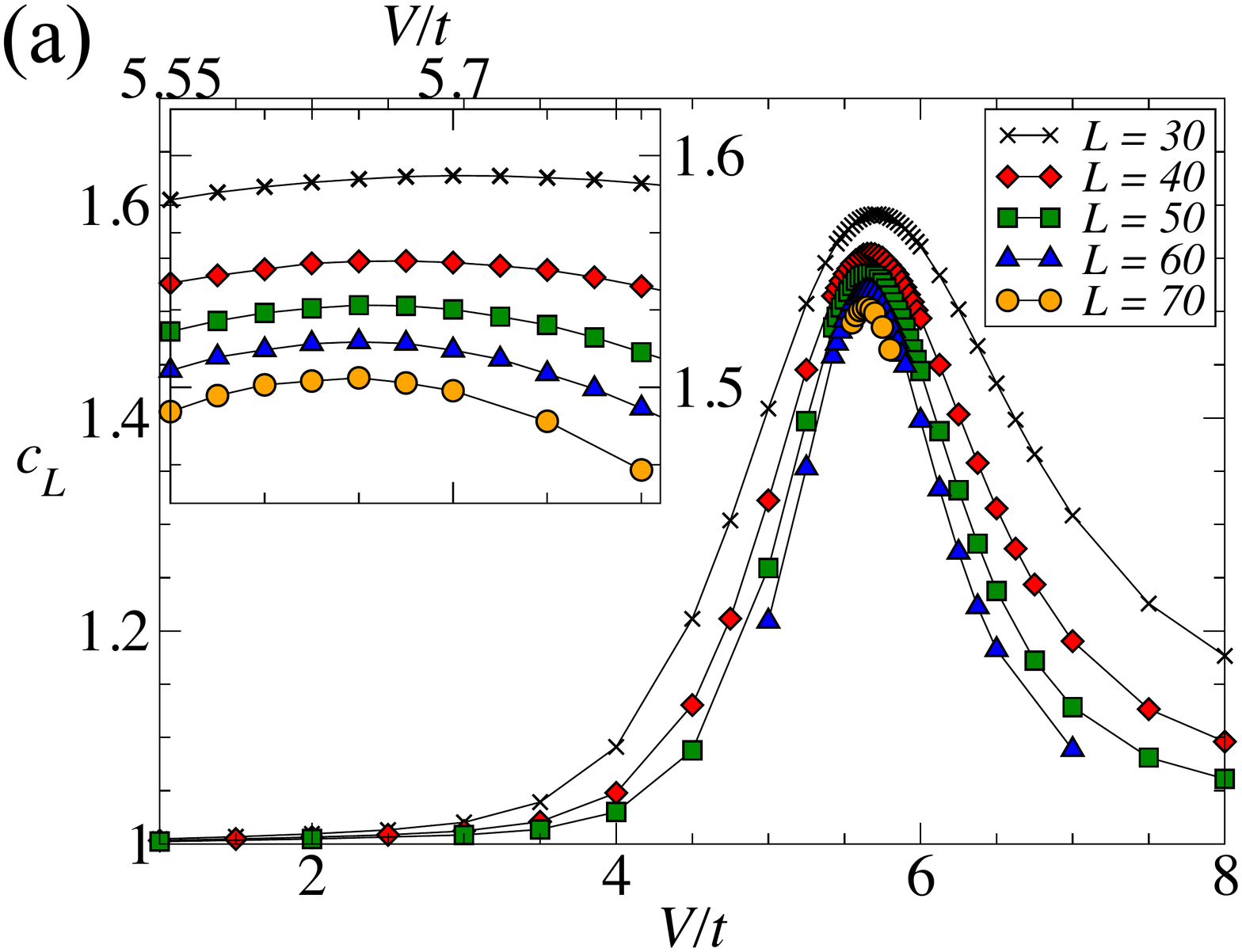}
\includegraphics[width = 0.85\columnwidth]{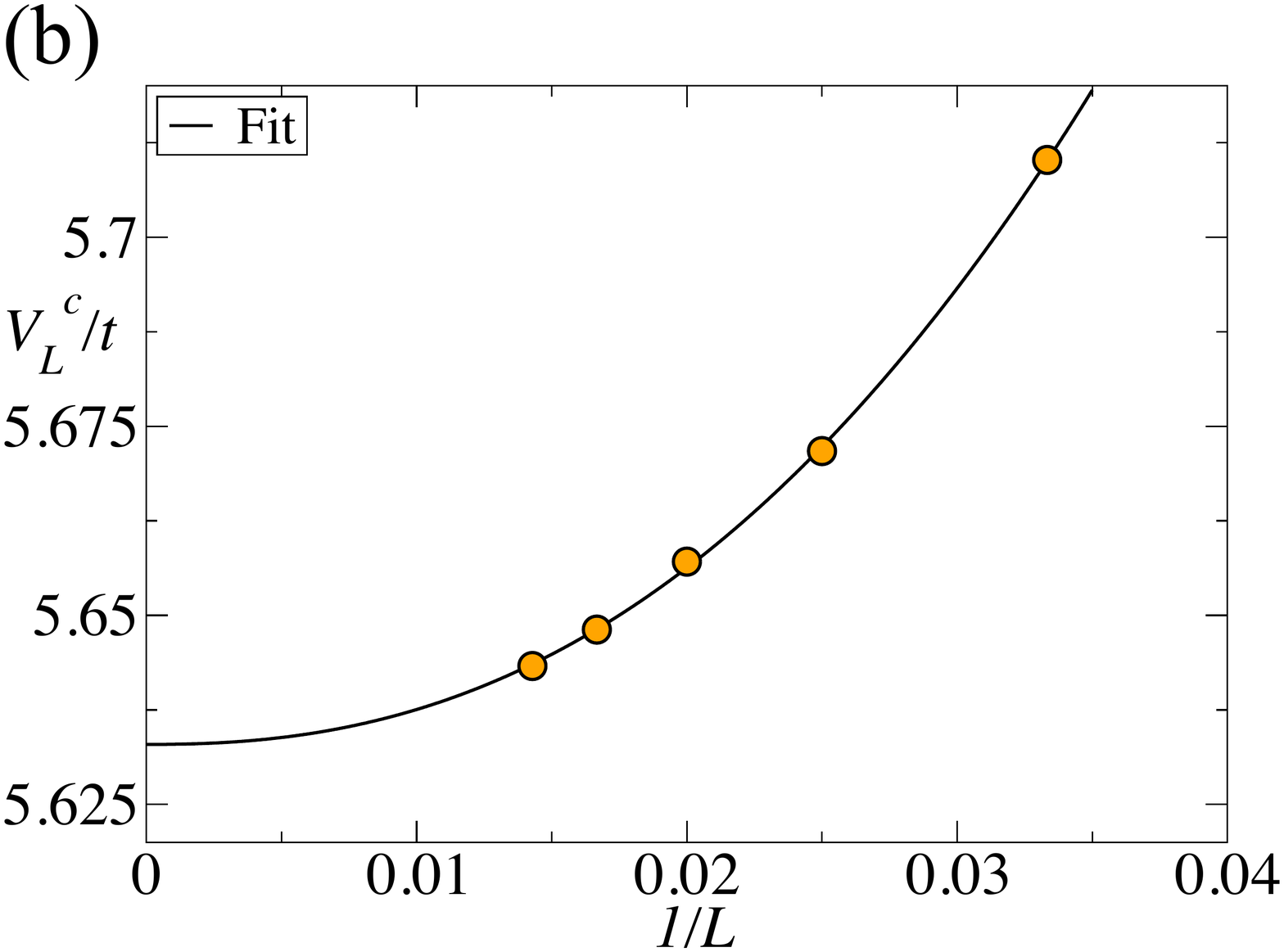}
\caption{Panel (a): Finite-size central charge as estimated from a linear interpolation according to the Cardy-Calabrese formula. 
Blocks of size $<9$ are neglected to avoid strong non-universal effects. Residuals are of order $10^{-4}$ in the worst
cases, closer to the transition line and at large system sizes. The data evidence a phase transition around $V/t\simeq 5.63\pm0.02$
with a $c=3/2$ central charge. Both liquid phases are compactified bosons, as expected. Panel (b): Finite-size scaling for the critical point $V_L^c$ as a function of $1/L$. Errors in the estimates of $V_L^c$, obtained via local interpolations of $c_L$ in panel (a), are of the size of the symbols. 
\label{fig_cc_V}}  
\end{center}
\end{figure}

A summary of all data is presented in Fig.~\ref{fig_cc_V}(a), where we show the dependence of $c(L)$ as a function of $V/t$ for different system sizes and $n=4/10$. The data strongly suggest that there is an intervening phase transition between the two different $c=1$ phases (TLL and CLL) at an intermediate value of $V/t$, evidenced by the bell-like structure centered around $V/t\simeq 5.65$.

At the transition point, the central charge is compatible with $c=3/2$, as shown in Fig.~\ref{fig_S}(b). This universality class points toward the presence of a supersymmetric critical point~\cite{Dix88}, where the low-energy field theory is described by a combination of an compactified boson, and a real (Majorana) fermion~\cite{difrancesco_book}. Critical points of this kind have been recently discussed in different situations, such as constrained models coming from explicitly supersymmetric Hamiltonians~\cite{Fen2003, Bauer} and as effective boundary degrees of freedom for topological phases~\cite{Gro14}.

In order to extract the transition point more accurately, in Fig.~\ref{fig_cc_V}(b) we plot as a function of the system size, the critical value of $V^c_L$ at which the finite-size central charge reaches its maximum value, and then take the thermodynamic limit. We obtain an estimate of the critical point $V^c/t\simeq 5.63\pm 0.02 $.

\subsection{Gaps and low-energy degrees of freedom} \label{subsec:gaps}

In order to further deepen our understanding of the two phases and of the low-energy excitations in the vicinity of the transition point, we analyze the lower part of the spectrum of the Hamiltonian. In particular, we have targeted the single particle gap, defined as
\begin{equation}
\Delta_{sp}(L)=2E_{N}(L)-E_{N-1}(L)-E_{N+1}(L),
\end{equation}
where $E_N(L)$ is the ground state energy in a chain of length $L$ with $N$ particles. In addition, we have investigated the {\it cluster gap} $\Delta_{cl}$, which, for the case $n=4/10$, is defined as
\begin{equation}
\Delta_{cl}(L) = 2E_{N}(L) - E_{N-2}(L) - E_{N+2}(L).
\end{equation}
In the TLL phase, both gaps are expected to vanish, as they are linear combinations of different vertex operators. In the cluster phase, however, the picture is different, as discussed below. In the following, we first   analyze the gaps in the classical limit $t=0$ by considering the exact solution of Sec.~\ref{sec:model_ham} and then present predictions of field theory and exact numerical results for $t\neq 0$.\\

\paragraph{Single-particle gap in a clustered state:}
Let us consider a system with size $L=10 \ell$, $N_A=2\ell$ and $N_B=\ell$, where $\ell $ labels the number of building blocks. The classical energy of the system is
\begin{equation}
E_N=V N_B = V\ell
\end{equation}
and, upon doping, one gets
\begin{equation}
E_{N+1} = V\ell +2V,\quad E_{N-1} = V\ell - V, 
\end{equation}
as the states with $N+1$ and $N-1$ particles cannot rearrange properly due to cluster constraints. This implies that the single particle gap is always open in the cluster phase, and in particular $\Delta_{sp}(L) = V$ for every system size. \\

\paragraph{Cluster gap in a clustered state:}
For the cluster gap the situation is remarkably different. Doping with a B cluster (two-particles) generates the following configuration in the classical limit
\begin{equation}
E_{N+2} = V\ell +3V,\quad E_{N-2} = V\ell - 3V 
\end{equation}
The latter is very much reminiscent of a classical crystal deformation - insertion and extraction of a single cluster take the same energy from the system. This leads to a vanishing cluster gap, as both contribution exactly cancel.\\

From the field theory described in Subsec.~\ref{subsec:lowe}, one can see that, while the cluster gap is nothing but a combination of vertex operators in the cluster language and thus vanishes in the cluster phase, the single particle gap cannot be written in this way, thus implying that single particle excitations are never gapless within the low-energy description. This implies that there must be a phase transition between the CLL and TLL phases where the single particle gap $\Delta_{sp}$ opens, while the cluster gap $\Delta_{cl}$ remains largely unperturbed.\\

\begin{figure}[t]
\begin{center}
\includegraphics[width = 0.48\columnwidth]{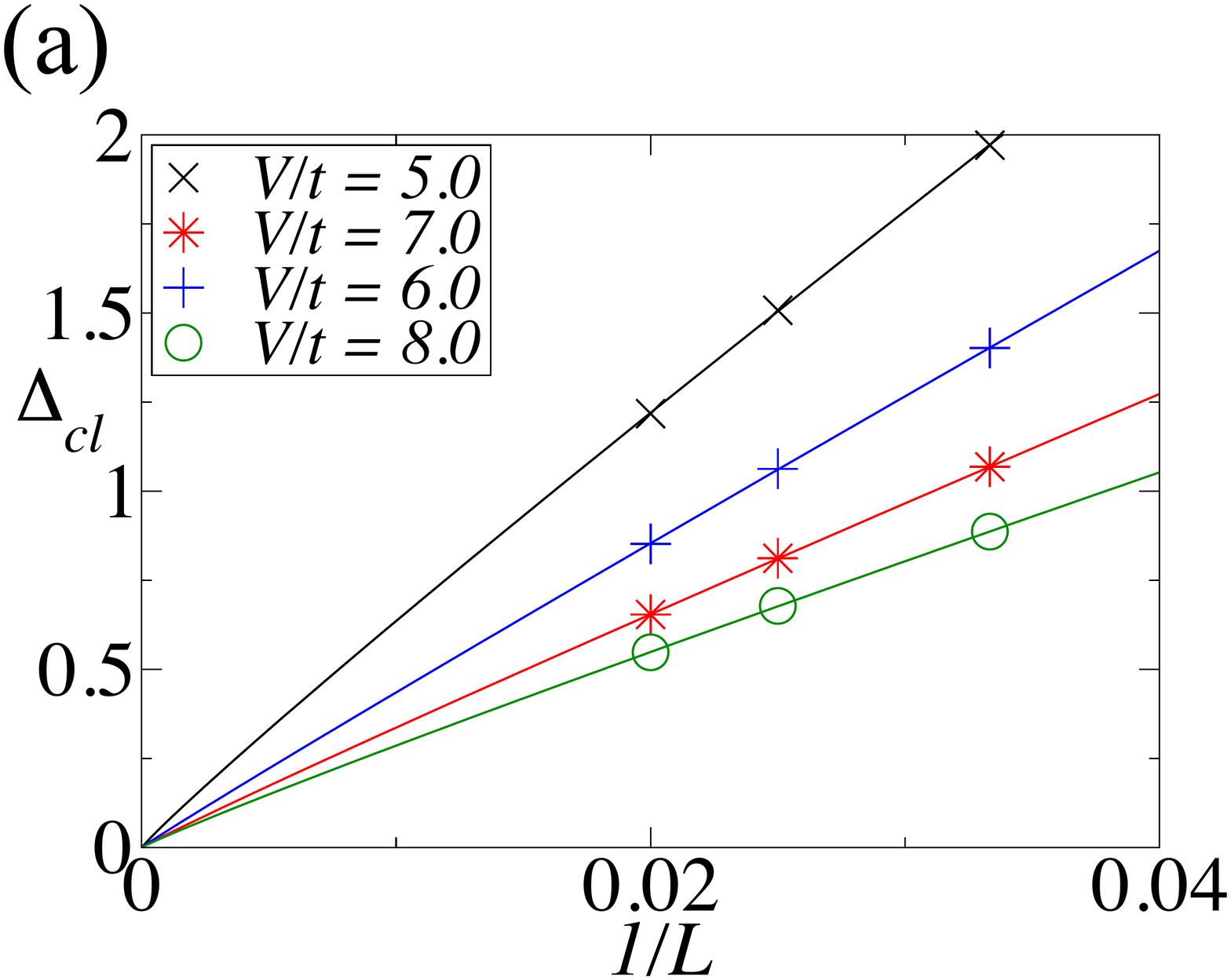}
\includegraphics[width = 0.48\columnwidth]{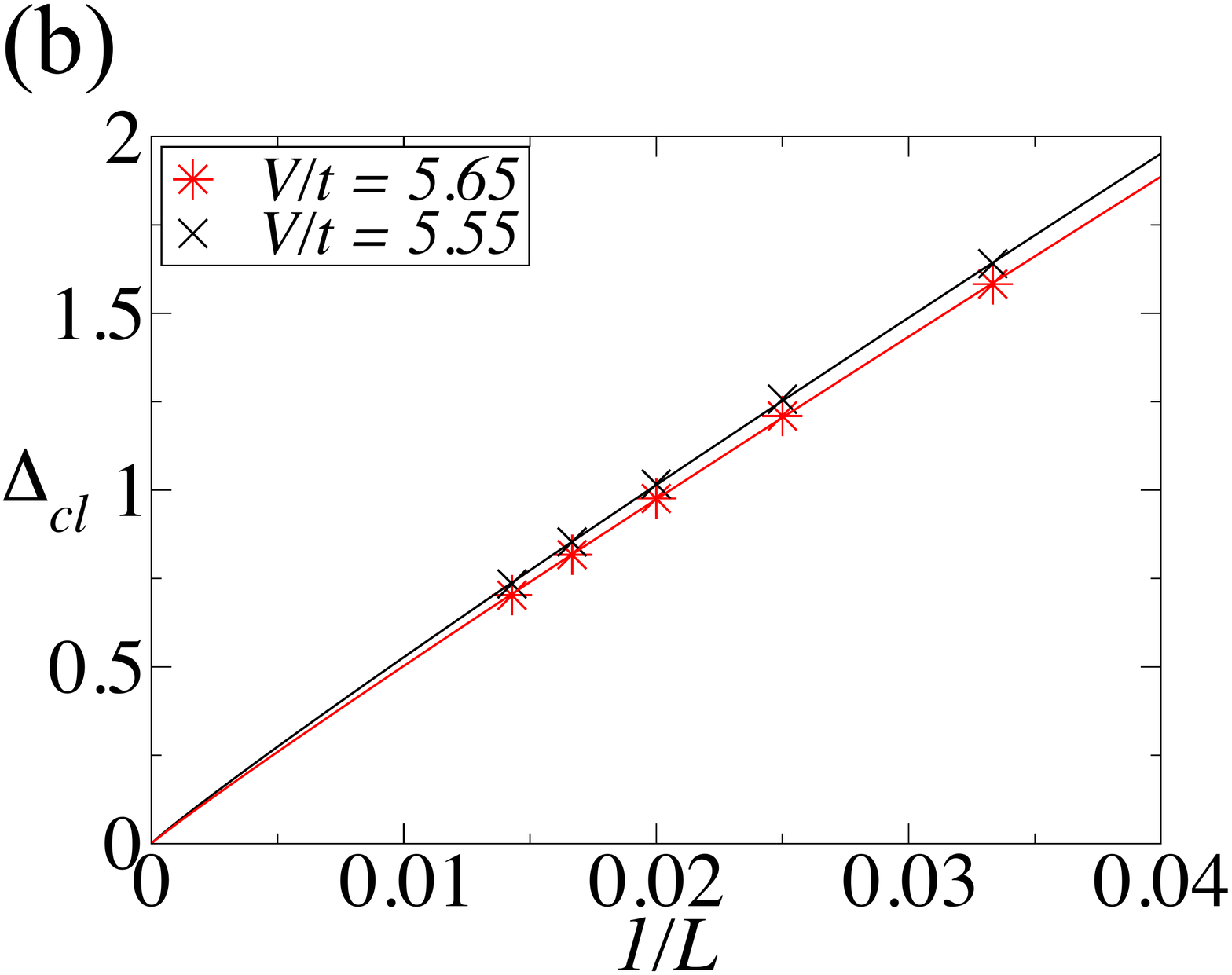}
\caption{Panel (a): Finite-size scaling of the cluster gap deep into the different phases. The gap scales to 0 in both CLL and TLL phases. Panel (b): data close to the transition point. 
\label{Cgap1}}  
\end{center}
\end{figure}

\begin{figure}[t]
\begin{center}
\includegraphics[width = 0.65\columnwidth]{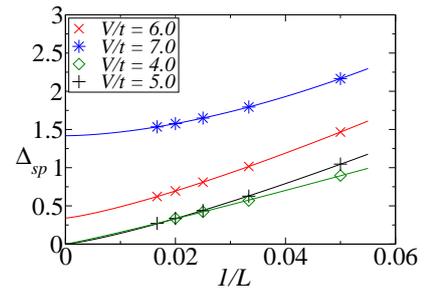}
\caption{Finite-size scaling of the single particle gap deep into the different phases. The gap scales to 0 in the TLL phase, but takes a finite expectation value in the CLL phase, in agreement with the field theory predictions. 
\label{Spgap}}  
\end{center}
\end{figure}
The numerical results on the gap scaling fully confirm this picture. In Fig.~\ref{Cgap1}(a), we show the cluster gap $\Delta_{cl}$ for some points representative of the TLL ($V/t = 5$) and cluster phase ($V/t=6,7,8$). In both phases, $\Delta_{cl}$ scales to 0 in the thermodynamic limit as a power law - as predicted from the field theory analysis. Moreover, the scaling is also present close to the transition point [Fig.~\ref{Cgap1}(b)], where the gap vanishes approximately as $\propto1/L$. 

\begin{figure}[t]
\begin{center}
\includegraphics[width = 0.65\columnwidth]{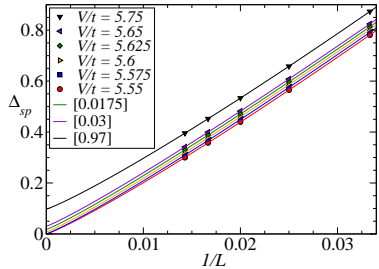}
\caption{Finite-size scaling of the single particle gap in the vicinity of the transition point. Lines are best fits of the form $a_0+ a_1(1/L^{a_2})$, with $a_0$, $a_1$ and $a_2$ constants; the caption indicates as a reference some of the extrapolated values. 
\label{Spgap2}}  
\end{center}
\end{figure}

\begin{figure}[t]
\begin{center}
\includegraphics[width = 0.65\columnwidth]{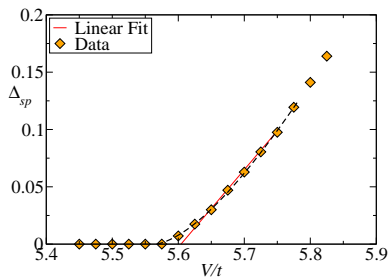}
\caption{Single particle gap extrapolated to the thermodynamic limit as a function of the interaction strength. The red line is a linear fit in the vicinity of the transition point, while the dashed line is a linear fit improved with a logarithmic correction. Errors in the extrapolation are of order of the symbol-size ($\simeq 0.01$) except at $V/t=5.6, 5.675$, where the errors estimated by least-square methods are $\simeq 0.015$.
\label{Spgap3}}  
\end{center}
\end{figure}

The behaviour of the single particle gap $\Delta_{sp}$ in the two phases is illustrated in Fig.~\ref{Spgap}. Deep in the TLL phase, the gap scales to 0 (green diamonds and black pluses). However, once in the CLL phase, the gap is clearly finite, and extrapolates to larger values as $V$ increases. We notice that in this latter region we have excluded from the fit to the data small system sizes with $L=10$, which might display strong finite-size effects due to the presence of limited cluster structures at very small sizes. 

In Figs.~\ref{Spgap2} and~\ref{Spgap3} we show how the single particle gap $\Delta_{sp}$ scales in the vicinity of the transition point.  In particular, Fig.~\ref{Spgap3} shows that the dependence of the gap on $V/t$ is approximately linear (red dashed line), which is consistent with an emergent Ising field at the critical point~\cite{Dix88}. The linear extrapolations locate the critical point around $V/t\simeq 5.605$, in good agreement with the results for the scaling of the entanglement entropy (see Subsec.~\ref{subsec:entr}). Moreover, we notice that the agreement improves considerably once additional logarithmic corrections are included in the fit. This could point towards the presence of additional corrections close to the critical point, which have already been observed in similar supersymmetric scenarios~\cite{Bauer}. An accurate understanding of the relation between the microscopic degrees of freedom and the emergent fields at the critical point might shed further light onto these finite-size corrections. 

\subsection{Sound velocities}

\begin{figure}[t]
\begin{center}
\includegraphics[width = 0.75\columnwidth]{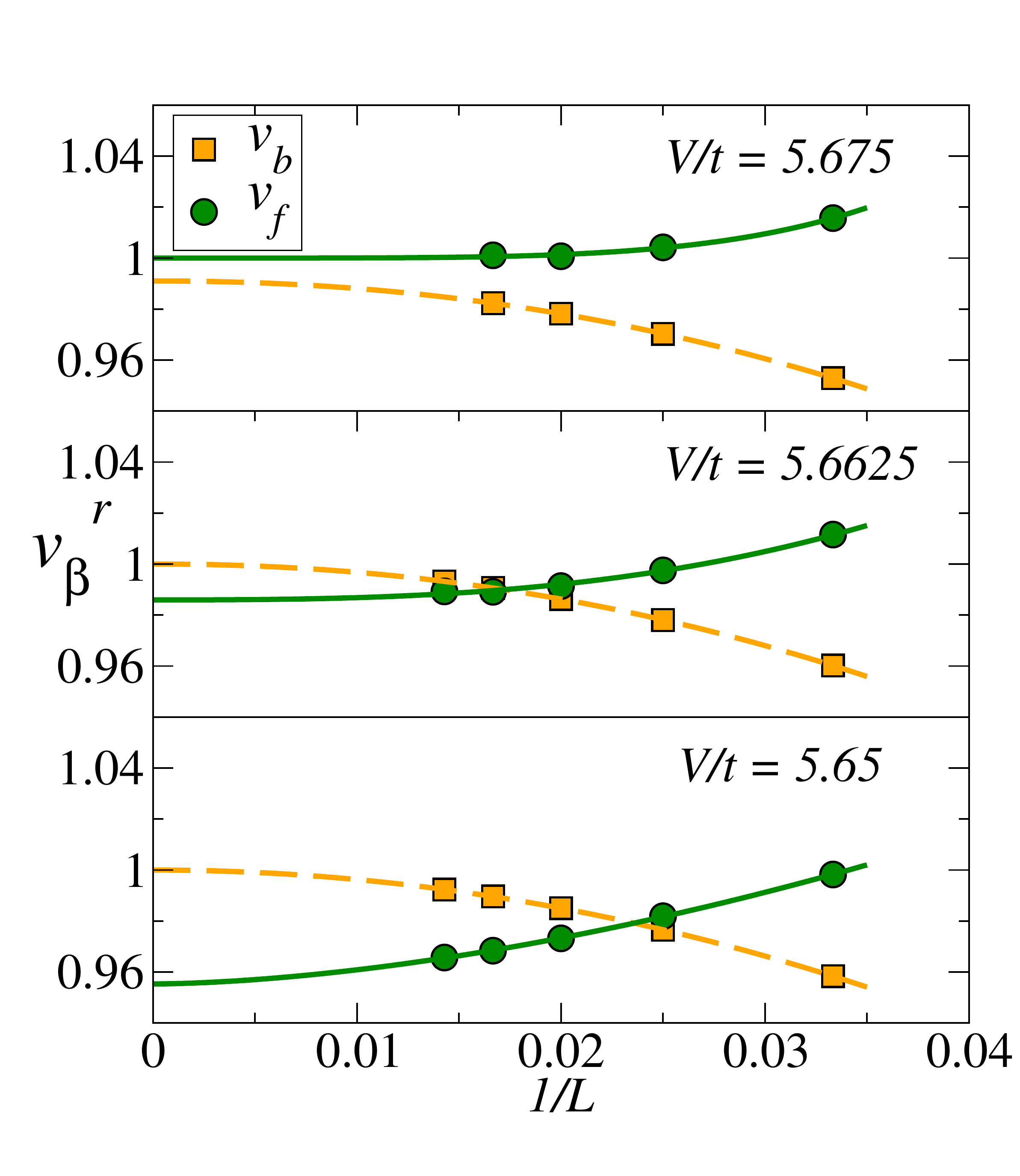}
\caption{Rescaled sound velocities $v_\beta^r = v^\beta/v_M$ (where $v_M$ is the maximum of the two sound velocity in the $L\rightarrow\infty$ limit) of the fermionic and bosonic models are extracted from the low-energy spectrum following Eq.~\ref{vb},~\ref{vc}.
\label{sound}}  
\end{center}
\end{figure}

At the critical point, the emergent supersymmetry implies that the scaling dimensions of the operators are fixed, and the sound velocities of both the boson and Ising (fermionic) mode are the same. Within our framework, we can identify the sound velocity of the bosonic mode $v_B$ using conventional conformal field theory techniques (see, e.g., Ref.~\onlinecite{henkel_book}), by monitoring the finite-size scaling of the cluster gap~\footnote{The cluster gap, as it is defined here, is driven by a pair of vertex operators, so technical speaking, the bosonic gap of the continuum theory is half the cluster gap.}:
\begin{equation}
\Delta_{cl}/2 = \frac{2\pi v_Bx_B}{L}
\end{equation}
where $x_B$ is the scaling dimension of the corresponding vertex operator. Following the scaling dimensions for the expected value of $x_B=1/4K$ at criticality, with $K=4$ being the Luttinger parameter~\cite{Bauer}, one gets:
\begin{equation}\label{vb}
v_B = \frac{4\Delta_{cl}L}{\pi}.
\end{equation}
The estimate of the fermionic velocity is however not unambiguous, as our field theory does not allow to establish an exact mapping between the low-energy, continuum fields, and the lattice operators. As the gap in the Ising model approaches $1$ in the strong coupling limit of the Ising model following the notations in Ref.~\onlinecite{henkel_book}, while our $\Delta_{sp}$ approaches $V$ and scales approximately as $\simeq(V-V_c)$ in the vicinity of the critical point, it seems plausible to assume that $\Delta_{sp} = \Delta_{F}$ in our context, where $\Delta_F$ is the gap in the Ising model corresponding to different parity sectors - thus related to the {\it spin} primary operator. This implies:
\begin{equation}\label{vc}
\Delta_{sp} = \frac{2\pi v_f( \bar{\delta}_\sigma + \delta_\sigma)}{L}
\end{equation}
where $\bar{\delta}_\sigma,\delta_\sigma = 1/16$ are the anti-holomorphic and holomorphic dimensions of the primary operator corresponding to the Ising spin at low energy. We thus get the following form for the fermionic sound velocity:
\begin{equation}
v_f = \frac{8\Delta_{sp}L}{\pi}.
\end{equation}
In Fig.~\ref{sound}, we report our results for the velocities in Eq.~\ref{vb},~\ref{vc} in the vicinity of the critical point. The $y$-axis are rescaled by the value of the largest velocity in the $L\rightarrow\infty$ limit to improve readability. The fits, indicated by the continuos lines, are 3-parameter fits of the form $a_0+a_1*L^{a_2}$, where the extrapolated value $a_0$ represents the velocity in the thermodynamic limit. Typical errors in the extrapolated limit are of order 3\%~\footnote{The error is estimated by performing the fits with a different set of points, and keeping track of the variation in the value of $a_0$. The term $a_1 * L^{a2}$ represents corrections beyond the linear order which have non-universal nature (see, e.g., the discussion in Ref.~\onlinecite{henkel_book}).}. As can be seen in the upper panel, the velocities are equal (within numerical error) in the close vicinity of the transition point, further supporting the supersymmetric nature of the critical point itself. 
We note that the next order corrections to the fermionic velocity are scaling with $a_2\simeq -2$, in full agreement with the conformal field theory prediction~\cite{henkel_book}.

%% file: cluster_liquid.tex
\section{Resilience of cluster features against temperature} \label{sec:temp}

In this section, we study the finite temperature properties of the cluster Luttinger liquid state. Even though thermal fluctuations will suppress any  long-range or quasi-long range order for the 1D system at finite temperature, we will show that certain cluster features characterized by a peak in the static structure factor for certain quasi-momenta can still survive for temperatures of the order of the interaction strength.  To study the finite temperature properties of the system, we use a numerically exact Quantum Monte Carlo (QMC) method with worm-type updates~\cite{Prokofev1998} implemented in Ref.~\cite{Pollet2007}. 

\begin{figure}[htb]
\includegraphics[width=0.98\linewidth]{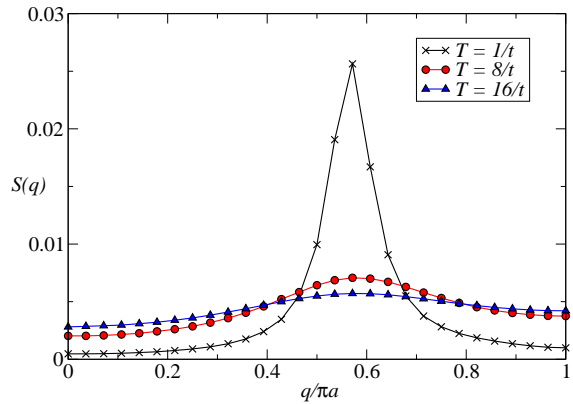}
\caption{The structure factor $S(q)$ for $V/t=8$, $L=56$. For example, $\mu=-3.2t$ $(-5t)$ for temperatures $T=1/t$ $(8/t)$,  such that $N\simeq 24$ is kept constant.}
\label{fig:finiteT}
\end{figure}

We choose the hard-core boson Hamiltonian we discussed above with $r_c=2$:
\begin{equation}
H=-t\sum_i (b_i^\dag b_{i+1}+h.c)+V\sum_i \sum_{\ell =1}^{r_c} n_i n_{i+\ell} - \sum_i \mu n_i \label{eq:hamT}.
\end{equation}
Here $\mu$ is the chemical potential, since our QMC simulations are performed in the grand canonical ensemble. We note that in Hamiltonian (\ref{eq:hamT}), $t>0$ and  the frustration induced by the next-nearest neighbour interactions only appears in diagonal terms. As a result, the system is free from the so-called "sign problem" irrespectively of the filling factor. In the QMC simulation, we choose the interaction strength $V=8t$ so that the ground state of the Hamiltonian corresponds to a cluster Luttinger liquid state. We focus on a chain with PBCs, and tune $\mu$ such that the average particle number is kept constant for the different temperatures studied.\\

In the following we focus on the static structure factor at finite temperature, in analogy to Eq.~\eqref{Sq}
\begin{equation}
S(q)=\frac{1}{L^2}\sum_{i,j} e^{iq(i-j)}\langle (n_i-n)(n_j-n)\rangle.
\end{equation}
Here $q$ is again the lattice quasi-momentum, $n=N/L$ is the average density and $\langle \hat{O}\rangle=Tr(\hat{O}e^{-\beta H})/Z$, with $Z$ the partition function at temperature $T=1/\beta$.
For a chain with $L=56$, $N=24$ and PBCs the number of blocks of type $A(B)$, $N_{A(B)}$, satisfy 
\begin{eqnarray}
N_A+2N_B= N\\
3N_A+4 N_B=L,
\end{eqnarray}
with $N_A=N_B=8$.
For these parameters, the cluster Luttinger liquid ground state exhibits a sharp peak in $S(q)$ at a characteristic wave vector $k_1= \pi (1-n)$ [from Eq.~\eqref{char_wv}], which indicates cluster features. In Fig.~\ref{fig:finiteT} we plot $S(q)$ and tune $\mu$ such that the average number of particles is $\langle N \rangle \simeq 24$, for the different temperatures studied ($T=1/t, 8/t$ and $16/t$). As expected, we find that the peaks, located at $k_1=4\pi/7$, become broader with increasing $T$. However, they remain clearly  visible up to temperatures of the order of $V$. This indicates that certain cluster features are comparatively robust against thermal fluctuations for sufficiently low temperatures.

%% file: adiabatic_1.tex
If a system is  prepared in the ground state of a certain initial Hamiltonian $H_0$ at time  $\tau=0$ and some parameter $\lambda(\tau)$ is adiabatically~\cite{landau, zener} tuned such that a different Hamiltonian $H_1$ is obtained at a final time $\tau=\tau_{\mathrm{max}}$, the adiabatic theorem~\cite{adiabatic_th} ensures that the original ground state will continuously evolve into the ground state of $H_1$. The total Hamiltonian $H(\tau)$ can be written as
\begin{equation}  \label{adprep1}
H(\tau) = [1-\lambda(\tau)] H_0 + \lambda(\tau) H_1,
\end{equation}
where $\lambda(\tau)$ ranges from $0$ to $1$ as time runs from $\tau=0$ to $\tau=\tau_{\mathrm{max}}$.

As a prototypical example, let's imagine to adiabatically prepare the CLL state of a system with $L=14$, $N=6$ and $r_c=2$, that is the ground state of $H_1 = -t_1\sum_i (b^\dagger_ib_{i+1}+\mathrm{H.c.})+V\sum_{i}\sum_{\ell=1}^{r_c}n_in_{i+\ell}$ for, e.g., $V/t_1 = 6$ (see Sec.~\ref{sec:phasd}). The initial Hamiltonian is $H_0=-t_0\sum_i (b^\dagger_ib_{i+1}+\mathrm{H.c.})+V\sum_{i}\sum_{\ell=1}^{r_c}n_in_{i+\ell}$, with $t_0 \ll t_1$.
Neglecting perturbatively small corrections $\propto t^2_0/V$ (see Subsec.~\ref{subsec:strong}), the ground state of $H_0$, under the assumption of PBC, is a mixed state of the 2-fold degenerate classical cluster configurations obtained from the possible permutations of two blocks of type $A$ and two of type $B$. For $L=14$, the allowed configurations are either $AABB$ or $ABAB$. \\

Since the interaction part in $H(\tau)$ is constant throughout the evolution, we can  rewrite Eq.~\eqref{adprep1} as
\begin{equation} \label{adprep2}
H(\tau) = \tilde{H}_0 + \beta(\tau) \tilde{H}_1,
\end{equation}
where $\tilde{H}_0=V\sum_{i}\sum_{\ell=1}^{r_c}n_in_{i+\ell}$, $\tilde{H}_1 = -\sum_i (b^\dagger_ib_{i+1}+\mathrm{H.c.})$ and $\beta(\tau) = t_0(1-\tau/\tau_{\mathrm{max}}) + t_1(\tau/\tau_{\mathrm{max}})$. Notice that, as in Eq.~\ref{adprep1}, $H(0) = H_0$ and $H(\tau_{\mathrm{max}}) = H_1$. \\

\begin{figure}[t]
\centerline{\includegraphics[width=0.9\columnwidth]{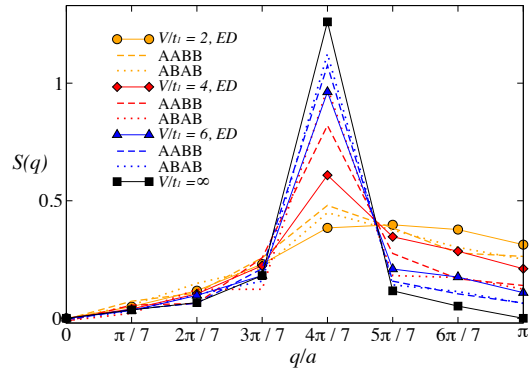}}
\caption{Comparison of the adiabatic state preparation at the final time $\tau_{\mathrm{max}}=200/t_1$ and starting from either $ABAB$ (dotted lines) or $AABB$ (dashed lines), with exact diagonalisation results (continuous lines plus symbols). Different colours distinguish different Hamiltonian parameters $V/t_1$. The ratio $t_0/t_1=0.01$ is fixed. Black squares indicate the classical limit prediction.}\label{fig:diffV}
\end{figure}

In Fig.~\ref{fig:diffV}, we show the static structure factor $S(q)$, at $\tau = \tau_{\mathrm{max}}$,
obtained by integrating the Schr\"odinger equation governed by the time-dependent Hamiltonian in Eq.~\eqref{adprep2} for different choices of $V/t_1$, while keeping $t_0/t_1$ fixed. We compare results of the adiabatic protocol with exact diagonalisation calculations of the final Hamiltonian $H_1$.

%---------------------------------------------------------------------
%---------------------------------------------------------------------
Since both $AABB$ and $ABAB$ {\it explicitly} break the aforementioned ground state degeneracy, the corresponding peaks of $S(q)$ have slightly different heights when compared to each other, despite being located at the same predicted critical momentum $k_1 = 4\pi/7$ of Eq.~\eqref{char_wv}. Independently of the choice of the initial state, we find that the best agreement between exact diagonalisation and  state preparation is obtained in the strongly interacting limit [blue lines in Fig.~\ref{fig:diffV}], where the CLL state is expected to be robust against quantum fluctuations (Sec.~\ref{sec:phasd}) and non-adiabatic effects. \\

Experimentally, the protocol we propose here can be implemented by starting with particles trapped in a very deep lattice, and then by reducing the optical lattice depth with a down-ramp that ensures a linear growth in time of $\beta(\tau)$. \\

In order to take into account possible losses due to off-resonant light scattering induced by the optical lattice and/or scattering of atoms with the background gas, we consider in addition a homogeneous and constant rate $\gamma$ of particle losses. We assume that $\gamma$ takes into account also non-homogenous losses originated by excitations to Rydberg states, which might experience different Stark shifts with respect to ground state atoms. Three-body collisions can be safely neglected due to the hard-core assumption $(b^{\dagger})^2 = 0$.  In the Born-Markov approximation, the reduced system density matrix $\rho(\tau)$ then evolves according to the master equation ($\hslash = 1$)
\begin{equation} \label{me}
\dot{\rho}(\tau) = -i\Big( H_{\mathrm{eff}}(\tau) \rho(\tau) - \rho(\tau) H^{\dagger}_{\mathrm{eff}}(\tau) \Big) +  \gamma \sum_{j=1}^L b_j \rho(\tau) b^{\dagger}_j,
\end{equation}
where $H_{\mathrm{eff}}(\tau) = H(\tau) - (i/2)\gamma \sum_{j=1}^L b^{\dagger}_j b_j$ is an effective non-Hermitian Hamiltonian and $H(\tau)$ is given by Eq.~(\ref{adprep2}). \\

\begin{figure}[ht]
\centerline{\includegraphics[width=0.9\columnwidth]{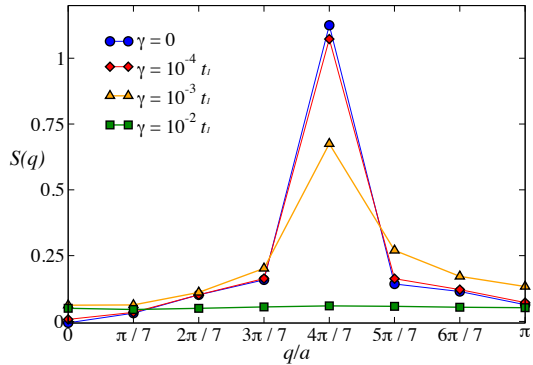}}
\caption{Plot of $S(q)$ for dissipative dynamics. The loss rates $\gamma$ range over three order of magnitudes. Also the case of purely Hamiltonian dynamics ($\gamma = 0$) is included. The initial state is $ABAB$, $\tau_{\mathrm{max}}= 200/t_1$ and $V/t_1 = 6$. Notice how cluster features survive at $\gamma = 10^{-3} t_1$.} \label{fig:diffg}
\end{figure}

We approximate the exact evolution in Eq.~\eqref{me} by Monte-Carlo trajectory simulations in the quantum jump approach~\cite{qj_dal,qj_zol, rev}. In Fig.~\ref{fig:diffg} we show $S(q)$ at $\tau=\tau_{\mathrm{max}}$, obtained by averaging over $N_{\mathrm{traj}} = 500$ different system realisations. As expected, the best approximation of the non-dissipative case is given by the smaller choice of $\gamma$. Nevertheless, the characteristic cluster peak in the static structure factor survives even for comparatively large loss rates, ensuring the stability of cluster features with respect to non-adiabatic effects in the protocol.

%---------------------------------------------------------------------
%---------------------------------------------------------------------

%% file: conclusions.tex
In summary, we have provided an in-depth study of the emergence of cluster Luttinger liquid phases in 1D models with soft-shoulder interactions. Starting from the exactly solvable classical limit, we have shown that different approaches such as strong coupling perturbation theory and constrained bosonization analysis predict the emergence of such a liquid phase of matter. In addition, we have analyzed numerically the full phase diagram for soft-shoulder potentials for different density regimes. Surprisingly, a critical point with central charge $c=3/2$ is found to separate the CLL and TLL phases, which have both $c=1$, indicating an emergent supersymmetry. While in the present work we have demonstrated that both the scaling of the entanglement entropy and of the gap provide evidence for such emergent behavior, and how the sound velocities of the emergent bosons and fermions are indeed equal within numerical accuracy, in future investigations it would be desirable to develop a field theoretical understanding of the emergent Ising field, and to provide a possible explanation of the stability of the critical point based on microscopic symmetries (as discussed in Ref.~\onlinecite{Bauer}). We remark that $c=3/2$ critical points have also been reported in bilinear-biquadratic spin-chains and multi-species Hubbard models in the presence of an explicit $\mathbb{Z}_2$ symmetry related to binding-unbinding mechanisms.~\cite{lecheminant2005,roux2009,manmana2011,ejima2011} \\

In order to address questions of experimental observability of the CLL phase in Rydberg-dressed gases, we have shown that finite temperature effects do not severely degrade the CLL signatures in correlation functions up to temperatures of the order of the interaction energy. We have further discussed how cluster states can  adiabatically be prepared on realistic time-scales where the effect of decoherence is largely non-detrimental. Since experiments have already been performed with atomic Rydberg chains comprising up to 40 sites, these studies suggest that the realization of the desired system dynamics may be accessible within state-of-the-art technology. In particular, the departure from the conventional Luttinger liquid scenario could be benchmarked by monitoring correlation functions either via noise-correlation measurements or by extracting  structure factors from single-site density measurements. \\

An intriguing extension for the models discussed here is the two-dimensional scenario. There, the classical model is also exactly solvable, albeit the final Hamiltonian looses the simple XY form derived in one-dimension. The possibility of emergent gauge fields and a violation of the Luttinger theorem in the fermionic case could lead to the stabilization of quantum spin liquid phases in a relatively simple model Hamiltonian, where cluster features play a prominent role.

%% file: cTL_MDM.bbl
\begin{thebibliography}{}



\bibitem{magnet_book} {\it Introduction to Frustrated Magnetism: Materials, Experiments, Theory}, edited by C. Lacroix, P. Mendels, and F. Mila (Springer, Heidelberg, 2011).
\bibitem{LIKOS2001} C. N. Likos, A. Lang, M. Watzlawek and H. L\"owen, Phys. Rev. E {\bf 63}, 031206 (2001).
\bibitem{MLADEK2006} B. M. Mladek {\it et al.} Phys. Rev. Lett. {\bf 96},  045701 (2006); C. N. Likos {\it et al.} J. Chem. Phys. {\bf 126}, 224502 (2007); B. M. Mladek {\it et al.}, Phys. Rev. Lett. {\bf 99}, 235702 (2007); J. Fornleitner and G. Kahl, J. Phys.: Condens. Matter {\bf 22}, 104118 (2010); D. Coslovich {\it et al.}, Soft Matter {\bf  7}, 2127 (2011). 
\bibitem{Boninsegni2012}M. Boninsegni and N. V. Prokof'ev, Rev. Mod. Phys. {\bf 84}, 759 (2012).
\bibitem{Cinti2010}F. Cinti, P. Jain, M. Boninsegni, A. Micheli, P. Zoller and G. Pupillo, Phys. Rev. Lett. {\bf 105}, 135301 (2010).
\bibitem{Henkel2010}N. Henkel, R. Nath, and T. Pohl, Phys. Rev. Lett. {\bf 104}, 195302 (2010).
\bibitem{Cinti2013}F. Cinti, T. Macr\`i, W. Lechner, G. Pupillo and T. Pohl, Nature Comm. {\bf 5}, 3235 (2014).
\bibitem{haldane1981}F. D. M. Haldane, Phys. Rev. Lett., \textbf{47}, 1840 (1981).
\bibitem{GIAMARCHI2003} T. Giamarchi, {\it Quantum Physics in one dimension}, (Oxford University press, Oxford, 2003).
\bibitem{gogolin_book}A. O. Gogolin, A. A. Nersesyan, and A. M. Tsvelik, \textit{Bosonization and strongly correlated systems}, (Cambridge University press, Cambridge, 1998), and references therein.
\bibitem{cazalilla2004} M. A. Cazalilla, J. Phys. B: At. Mol. Opt. Phys. {\bf 37}, 7 (2004).
\bibitem{cazalilla2011}M. A. Cazalilla, R. Citro, T. Giamarchi, E. Orignac and M. Rigol, Rev. Mod. Phys. {\bf 83}, 1405 (2011).
\bibitem{testll}T. Giamarchi, Int. J. Mod. Phys. B {\bf 26}, 1244004 (2012).
\bibitem{difrancesco_book}P. Di Francesco, P. Mathieu, and D. S\'{e}n\'{e}chal, {\it Conformal Field Theory}  (Springer-Verlag, New York, 1997).
\bibitem{hall}A. M. Chang, Rev. Mod. Phys. {\bf 75}, 1449 (2003), A. M. Chang, L. N. Pfeiffer, and K. W. West, Phys. Rev.  Lett. {\bf 77}, 2538 (1996).  
\bibitem{carbon}M. Bockrath, D. H. Cobden, J. Lu, A. G. Rinzler, R. E. Smalley, L. Balents, and P. L.
Mceuen, Nature {\bf 397}, 598-601 (1999).
\bibitem{Ion}J. W. Britton {\it et al.}, Nature {\bf 404}, 489-492 (2012); 
R. Islam {\it et al.}, Science {\bf 340}, 583-587 (2013); 
C. Schneider {\it et al.}, Rep. Prog. Phys. {\bf 75}, 024401 (2012).
\bibitem{Atom}K. Aikawa, A. Frisch, M. Mark, S. Baier, A. Rietzler, R. Grimm, and F. Ferlaino, Phys. Rev. Lett., \textbf{108}, 210401 (2012);
M. Lu, N. Q. Burdick, and B. L. Lev, Phys. Rev. Lett. {\bf 108}, 215301 (2012).
\bibitem{Mol1}A. Chotia, B. Neyenhuis, S. A. Moses, B. Yan, J. P.
Covey, M. Foss-Feig, A. M. Rey, D. S. Jin, and J. Ye, Phys.
Rev. Lett. {\bf 108}, 080405 (2012).
\bibitem{Mol2}T. Takekoshi {\it et al.}, Phys. Rev.  A {\bf 85}, 032506 (2012).
\bibitem{PRL_MMMDWLGP} M. Mattioli, M. Dalmonte, W. Lechner and G. Pupillo, Phys. Rev. Lett. {\bf 111}, 165302 (2013).
\bibitem{Fisher2013}H.-C. Jiang, M. S. Block, R. V. Mishmash, J. R. Garrison,
D. N. Sheng, O. I. Motrunich, and M. P. A. Fisher, Nature {\bf 493}, 39 (2013).
\bibitem{block2011}M. S. Block, R. V. Mishmash, R. K. Kaul, D. N. Sheng, O. I. Motrunich, and M. P. A. Fisher, Phys. Rev. Lett. {\bf 106}, 046402 (2011).
\bibitem{Mish2011}R. V. Mishmash, M. S. Block, R. K. Kaul, D. N. Sheng, O. I. Motrunich, M. P. A. Fisher, Phys. Rev. B {\bf 84}, 245127 (2011).
\bibitem{Sud2009}J. Sudan, A. Luscher, A. Laeuchli, Phys. Rev. B {\bf 80}, 140402 R (2009).
\bibitem{Pupillo2010}G. Pupillo, A. Micheli, M. Boninsegni, I. Lesanovsky, and P. Zoller, Phys. Rev. Lett. {\bf 104}, 223002 (2010).
\bibitem{Pohl2014}  T. Macr\'i and T. Pohl, Phys. Rev. A {\bf 89}, 011402(R) (2014). 
\bibitem{White}S. R. White, Phys. Rev. Lett. {\bf 69}, 2863 (1992).
\bibitem{Sch}U. Schollw\"ock, Rev. Mod. Phys. {\bf 77}, 259 (2005).
\bibitem{Dix88}L. Dixon, P. Ginsparg, and J. Harvey, Nucl. Phys. B {\bf306}, 470 (1988).
\bibitem{Fen2003}P. Fendley, K. Schoutens, and J. de Boer, Phys. Rev. Lett., {\bf90}, 120402 (2003).
\bibitem{Bauer}B. Bauer, L. Huijse, E. Berg, M. Troyer, and K. Schoutens, Phys. Rev. B {\bf 87}, 165145 (2013). 
\bibitem{revexp}D. Comparat and P. Pillet, J. Opt. Soc. Am. B {\bf 27}, A208 (2010); R. L\"ow, H. Weimer, J. Nipper, J. B. Balewski, B. Butscher, H. P. B\"uchler, T. Pfau, J. Phys. B {\bf 45}, 113001 (2012).
\bibitem{exp} J. D. Pritchard {\it et al.},  Phys. Rev. Lett. {\bf 105}, 193603 (2010); M. Viteau {\it et al.}, Phys. Rev. Lett. {\bf 107}, 060402 (2011); T. Peyronel {\it et al.},  Nature {\bf 488}, 57 (2012); P. Schau{\ss}  {\it et al.}, Nature {\bf 491}, 87 (2012); M. Viteau {\it et al.}, Phys. Rev. Lett. {\bf 109}, 053002 (2012); M. Robert-de-Saint-Vincent {\it et al.}, Phys. Rev. Lett. {\bf 110}, 045004 (2013); T. Baluktsian {\it et al.}, Phys. Rev. Lett. {\bf 110}, 123001 (2013).
\bibitem{Gallagher}T. F. Gallagher, {\em Rydberg Atoms} (Cambridge University Press, New York, 1994).
\bibitem{Santos2001}L. Santos {\it et al.}, Phys. Rev. Lett. {\bf 85}, 1791 (2000); J. Honer {\it et al.}, Phys. Rev. Lett. {\bf 105}, 160404 (2010).
\bibitem{Saffman2010}M. Saffman, T. G. Walker and K. M\o{}lmer, Rev. Mod. Phys. {\bf 82}, 2313 (2010).
\bibitem{Lahaye2009}T. Lahaye, C. Menotti, L. Santos, M. Lewenstein and T. Pfau, Rep. Prog. Phys. {\bf 72}, 126401 (2009).
\bibitem{Bloch2008}I. Bloch, J. Dalibard, and W. Zwerger, Rev. Mod. Phys. {\bf 80}, 885–964 (2008).
\bibitem{Baranov2012}M. A. Baranov, M. Dalmonte, G. Pupillo and P. Zoller,  Chem. Rev. {\bf 112}, 5012 (2012).
\bibitem{qj_dal} J. Dalibard, I. Castin, and K. M{\o}lmer,  Phys. Rev. Lett. \textbf{68}, 580 (1992).
\bibitem{qj_zol} R. Dum, P. Zoller, and H. Ritsch,  Phys. Rev. A  \textbf{45}, 4879 (1992).
\bibitem{rev} M. B. Plenio and P. L. Knight,  Rev. Mod. Phys. \textbf{70}, 101 (1998).
\bibitem{Zhuravlev2000}A. K. Zhuravlev and M. I. Katsnelson, Phys. Rev. B {\bf 64}, 033102 (2001).
\bibitem{Schmitteckert2004}P. Schmitteckert and R. Werner, Phys. Rev. B {\bf 69}, 195115 (2004).
\bibitem{Wil}C. Holzhey, F. Larsen, and F. Wilczek, Nucl. Phys. B {\bf 424}, 443 (1994).
\bibitem{calabrese}P. Calabrese and J. Cardy, J. Stat. Mech. {\bf 2004}, P06002 (2004).
\bibitem{Gro14}T. Grover, D. N. Sheng, and A. Vishwanath,  Science {\bf 344}, 280 (2014).
\bibitem{henkel_book} M. Henkel, {\it Conformal Invariance and Critical Phenomena}, Springer (1999).
\bibitem{Prokofev1998}
N. V. Prokof'ev, B. V. Svistunov, and I. S. Tupitsyn,
Phys. Lett. A {\bf238}, 253 (1998).
\bibitem{Pollet2007}
L. Pollet, K. V. Houcke, and S. M. A. Rombouts, J. Comp.
Phys {\bf225}, 2249 (2007).
\bibitem{landau} L. Landau, Phys. Zeit. Sow.  \textbf{2}, 46 (1932).
\bibitem{zener} C. Zener, Proc. R. Soc. Lond. \textbf{2}, 137 (1932).
\bibitem{adiabatic_th} M. Born and V. Fock, Zeit. f. Phys.  \textbf{51}, 165 (1928).
\bibitem{lecheminant2005}P. Lecheminant, E. Boulat, and P. Azaria, Phys. Rev. Lett. {\bf95}, 240402 (2005).
\bibitem{roux2009}G. Roux {\it et al.}, Eur. Phys. J. B {\bf68}, 293-308 (2009).
\bibitem{manmana2011}S. R. Manmana {\it et al.}, Phys. Rev. B {\bf83}, 184433 (2011).
\bibitem{ejima2011}S. Ejima {\it et al.}, Phys. Rev. Lett. {\bf106}, 015303 (2011).
\bibitem{qutip} J. R. Johansson, P. D. Nation, and F. Nori, Comp. Phys. Comm. {\bf 184}, 1234 (2013).
\end{thebibliography}
